\documentclass[3p]{elsarticle}

\usepackage{subfigure}
\usepackage{verbatim}
\usepackage{amsmath}
\usepackage{amsfonts}
\usepackage{amssymb}
\usepackage{graphicx}
\usepackage{lscape}
\usepackage{multicol}
\usepackage{centernot}
\usepackage{floatpag}

\usepackage{adjustbox}
\usepackage{array}
\usepackage{booktabs}
\usepackage{multirow}

\usepackage{colortbl}
\definecolor{Grey9}{rgb}{0.9,0.9,0.9}
\definecolor{Grey7}{rgb}{0.7,0.7,0.7}
\newcolumntype{R}[2]{%
    >{\adjustbox{angle=#1,lap=\width-(#2)}\bgroup}%
    l%
    <{\egroup}%
}

\journal{Internet of Things}

\begin{document}

\begin{frontmatter}

\title{The Internet of Things, Fog and Cloud Continuum: Integration and Challenges}
%\title{From Things to the Cloud Through the Fog: Concepts, Review, and Challenges}
%\title{Internet of Things, Fog Computing and Cloud Hierarchy: Concepts, Review, and Challenges}
%\title{Internet of Things to the Cloud through the Fog: Concepts, Management, and Challenges}
%\title{Internet of Things, Fog Computing and Cloud Hierarchy: Survey of Concepts and Challenges}

% or include affiliations in footnotes:
 \author[unicamp]{Luiz Bittencourt}
 \ead{bit@ic.unicamp.br}
 \author[unicamp]{Roger Immich}
 \ead{roger@ic.unicamp.br}
 \author[uman]{Rizos Sakellariou}
 \ead{rizos@manchester.ac.uk}
 \author[unicamp]{Nelson Fonseca}
 \ead{nfonseca@ic.unicamp.br}
 \author[unicamp]{Edmundo Madeira}
 \ead{edmundo@ic.unicamp.br}
 \author[coimbra]{Marilia Curado}
 \ead{marilia@dei.uc.pt}
 \author[unicamp]{Leandro Villas}
 \ead{leandro@ic.unicamp.br}
 \author[tcd]{Luiz da Silva}
 \ead{dasilval@tcd.ie}
 \author[aero]{Craig Lee}
 \ead{lee@aero.org}
 \author[cardiff]{Omer Rana}
 \ead{ranaof@cardiff.ac.uk}

 \address[unicamp]{Institute of Computing, University of Campinas, Brazil}
 \address[uman]{School of Computer Science, University of Manchester, UK}
 \address[coimbra]{Department of Informatics Engineering, University of Coimbra, Portugal}
 \address[tcd]{CONNECT Centre, Trinity College Dublin, Ireland}
 \address[aero]{The Aerospace Corporation, USA}
 \address[cardiff]{School of Computer Science, Cardiff University, UK}

\begin{abstract}
The	Internet of Things needs for computing power and storage are expected to remain on the rise in the next decade. Consequently, the amount of data generated by devices at the edge of the network will also grow. While cloud computing has been an established and effective way of acquiring computation and storage as a service to many applications, it may not be suitable to handle the myriad of data from IoT devices and fulfill largely heterogeneous application requirements. Fog computing has been developed to lie between IoT and the cloud, providing a hierarchy of computing power that can collect, aggregate, and process data from/to IoT devices. Combining fog and cloud may reduce data transfers and communication bottlenecks to the cloud and also contribute to reduced latencies, as fog computing resources exist closer to the edge. This paper examines this IoT-Fog-Cloud ecosystem and provides a literature review from different facets of it: how it can be organized, how management is being addressed, and how applications can benefit from it. Lastly, we present challenging issues yet to be addressed in IoT-Fog-Cloud infrastructures.
\end{abstract}


\end{frontmatter}

%\tableofcontents

%\linenumbers

%%%%%%%%%%%%%
%  Introduction %%%%%
%%%%%%%%%%%%%
\section{Introduction} 
    Electronic devices have been expanding their ability to generate data, resulting in the accumulation of a wide variety of information ranging from measurements of natural phenomena to human-related behavior. With the expected expansion of the Internet of Things~(IoT)~\cite{GUBBI20131645}, it is predicted that in the future virtually all objects will be in some way connected. Along with the connection of {\em everything} to the Internet comes the need for transferring, storing, and processing unprecedented amounts of data, laying down a path for many years of research related to such IoT requirements.
    
    Cloud computing has evolved and became an easy-to-use platform for applications in general to store and process data. IoT devices frequently rely on cloud computing to store and process data, producing information and knowledge as a result. On the one hand, the wide adoption of cloud computing is a consequence of a fast time-to-market for many types of applications due to the paradigm's flexibility and reduced or null initial capital expenditures. %(CAPEX). 
    On the other hand, this same wide adoption has exposed some limitations of the paradigm in fulfilling all requirements of some classes of applications, such as real-time, low latency, and mobile applications. The centralized cloud data centers are often physically and/or logically distant from the cloud client, implying communication and data transfers to traverse multiple hops, which introduces delays and consumes network bandwidth of edge and core networks~\cite{7912261}.
    
    The widespread adoption of cloud computing, combined with the ever increasing ability of edge devices to run heterogeneous applications that generate and consume all kinds of data from a variety of sources, requires novel distributed computing infrastructures that can cope with such heterogeneous application requirements. Computing infrastructures that enact applications at edge devices have started to appear in recent years~\cite{dinh2013survey,shi2016edge}, improving aspects such as response time and reducing bandwidth use. Combining the ability of running smaller, localized applications at the edge with the high-capacity from the cloud, fog computing has emerged as an paradigm that can support heterogeneous requirements of small and large applications through multiple layers of a computational infrastructure that combines resources from the edge of the network as well as from the cloud~\cite{Bonomi:2012:FCR:2342509.2342513}.
    
    In this paper, we aim at identifying and reviewing the main aspects and challenges that make the combination of fog computing and cloud computing suitable for all kinds of applications leveraged by the Internet of Things. We discuss aspects from the infrastructure (processing, networking, protocols, and infrastructure for 5G support) to applications (smart cities, urban computing, and industry 4.0), passing through the management complexity of the distributed IoT-fog-cloud system (services, resource allocation and optimization, energy consumption, data management and locality, devices federation and trust, and business and service models). 
    
    In the next section we introduce concepts and definitions for Internet of Things (IoT), cloud computing, and fog computing. In Section \ref{sec:review} we review the literature on the aforementioned infrastructure, management, and applications aspects. Section \ref{sec:challenges} discusses several challenges related to these aspects, and Section \ref{sec:con} presents concluding remarks.

%%%%%%%%%%%%%%%%%%%%%
% Cloud, Fog, and IoT: basic definitions %
%%%%%%%%%%%%%%%%%%%%%
\section{IoT, Fog, and Cloud: Basic Definitions}
%Try to cover all definitions necessary to the literature review and challenges
    This section introduces the terminology and concepts related to the three components of the IoT-Fog-Cloud ecosystem. 

\subsection{Internet of Things}  
    Several predictions about the Internet of Things have been put forward in the literature. If one consensus exists about IoT, it is about the number of connected devices: dozens of billions of ``things’’ will be connected in a few years from now~\cite{GUBBI20131645,sundmaeker2010vision}. Such devices can include virtually any object with embedded microcontroller and communication capabilities (e.g., in a generalized manner, a set of sensors and/or actuators).
    
    This unprecedented number of devices results in an unprecedented amount of data to be transmitted and processed. More than that, IoT connected devices are highly heterogeneous at many levels: data communication protocols, energy requirements, computing capacity, mobility, and so on. Therefore, IoT devices management, throughout the data communication and processing stack, becomes intrinsically challenging.
    
    Raw data generated by the Internet of Things as a whole may not be directly useful. Such extraordinarily large data sets require significant processing and knowledge extraction capabilities to provide some insightful information.
    % visualization of such extraordinarily large data sets. 
    IoT applications are aimed at realizing this task: transform gathered data into actual information knowledge. Although a myriad of new applications is enabled by IoT, this is also a source of increased heterogeneity: different applications also have different requirements, which should be fulfilled by the computing system amalgamating IoT devices with their applications. In the next sections, we present two computing paradigms that can be utilized together to fulfill the heterogeneous requirements associated with IoT applications: cloud and fog computing.

\subsection{Cloud computing}
    Cloud computing has achieved a mature state in the past decade, turning into a widely adopted computing paradigm for many applications, due to its dynamic characteristics such as elasticity and pay-per-use. To be able to provide these characteristics, virtualization is one of the management pillars for cloud providers. Virtual machines and containers allow providers to share slices of their computing resources, usually deployed in large data centers, among users, resulting in a logically isolated system for each tenant.
    
    On-demand computing is offered by cloud providers based on three canonical models, namely Infrastructure as a Service (IaaS), Platform as a Service (PaaS), and Software as a Service (SaaS)~\cite{Armbrust:2010:VCC:1721654.1721672}. IaaS offers computing infrastructure as a service, where the user can remotely access and manage computing power; PaaS offers a platform for software development along with the necessary libraries and databases to deploy and run applications, and SaaS offers the software itself relying on the cloud provider’s infrastructure to offload computing and/or data. A variety of cloud service levels has surfaced, resulting in the Everything as a Service (XaaS) concept~\cite{7214098}.
    
    Cloud providers can also be classified according to their deployment model: public, private, hybrid and community clouds. Public clouds are those open to the public, usually charged on a pay-per-use basis for anyone with an Internet connection. Private clouds are restricted to a set of predefined users (e.g., from a company or university). Hybrid clouds are a composition of public and private cloud resources, often composed to fulfill the dynamic demand and avoid upfront investment for peak demand~\cite{6295710}. The community clouds~\cite{mell2011nist}, resembling the virtual organizations from grid computing\cite{foster2001anatomy},  are a composition of private clouds in order to share resources.
    
    Cloud services are offered based on a Service Level Agreement (SLA), which establishes what is offered and how the user should be charged to use the cloud service. Common examples are pay-per-use models where charging takes place by time unit (e.g., virtual machines in a per hour basis), by amount of data (e.g., data transfers off the provider or amount of data stored), or by number of requests (e.g., the number of times a specific function/method was called in the programming model of a SaaS provider).
    
    The above characteristics result in properties that make the cloud attractive for clients, as for example on-demand provisioning/deprovisioning, elasticity, ubiquitous access, lower upfront investments with reduced capital expenditures %(CAPEX) 
in exchange for operational expenditures, %(OPEX), 
and faster time to market. Throughout this paper, we discuss how clouds can fulfill part of the application requirements within the IoT landscape. We also discuss how fog computing, defined in the next section, can be combined with the cloud to provide an infrastructure that fulfills a wide range of requirements for IoT applications.
    
\subsection{Fog Computing}
    Computing capacity at the edge increased with the hardware evolution of personal devices. The combination of higher computing capacity with the concomitant evolution of edge networks leveraged distributed computing paradigms that propose the utilization of edge devices to run applications and store data. The hardware evolution also allowed devices to shrink in size, resulting in mobile devices that have enough computing and battery capacity to run applications with reasonable complexity and quality of service (QoS).
    
    The aggregation of edge devices into a distributed system infrastructure has different names in the literature, also showing different characteristics and focuses. 
For example, the European Telecommunications Standards Institute~(ETSI) recently changed the name of Mobile Edge Computing to Multi-access edge computing, while keeping the same MEC acronym~\cite{ETSIMagazine2017}.
This change is an effort to provide a more flexible framework which goes beyond the new cellular operator's requirements.
The primary goal is still the same, that is, to provide cloud-like features close to the subscribers at the edge of the network, however, it now includes all other wired and wireless communication technologies.
Because of that, a broad range of new designs can be implemented at, for example, IoT and Vehicle-to-everything~(V2X) networks.
    
Fog computing brings together the edge devices and the cloud, as well as introduces a hierarchy of computing capacity (fog nodes, cloudlets or micro data centers) between the edge and the cloud~\cite{openfog2017openfog}. This capacity can be scattered at access points, routing devices in the network, the network core, and so on. It is expected that the higher in the network hierarchy a fog node (cloudlet or micro data center) is, the larger its computing capacity since it should provide capacity for a larger set of users downwards the hierarchy. Moreover, the lower in the hierarchy a device is, the closer to the edge it is, thus presenting lower communication delays to edge devices (e.g., end-user devices, IoT sensors and actuators, vehicles, drones, and so on).
    
    The computing hierarchy in the fog infrastructure can offer a wider range of service levels, supporting applications that cannot be supported by cloud computing alone. A fog infrastructure is able to handle applications with a variety of QoS requirements, as applications can run at a hierarchy level that provides adequate processing capacity and meets latency requirements. Another consequence of the use of processing closer to the edge is to reduce (aggregate) bandwidth use in the network along the path between edge and cloud.

%%%%%%%%%%%%%
% Literature review %%%
%%%%%%%%%%%%%
\section{Literature review}
\label{sec:review}
\begin{figure}[!htbp]
\centering
        \includegraphics[width=1.0\textwidth]{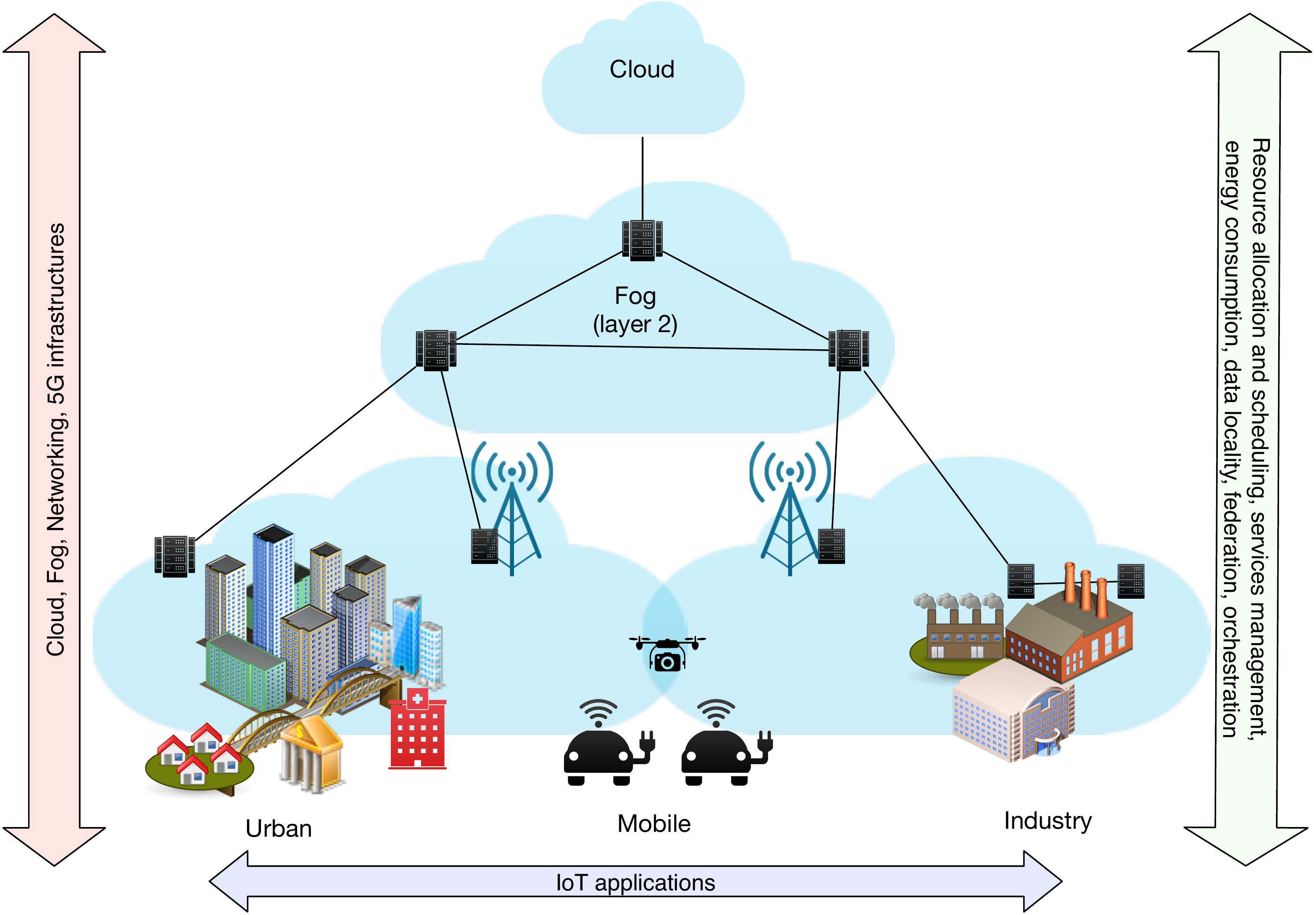}
\caption{Illustrative overview, within the IoT-Fog-Cloud infrastructure, of topics covered in this paper.}\label{fig:facets}
\end{figure}

In this section we discuss and review three different facets of the IoT-Fog-Cloud hierarchy, as illustrated in Figure~\ref{fig:facets}:

\begin{enumerate}
 \item \textbf{Infrastructure}, where the computing and networking infrastructure of cloud and fog is defined and discussed, including networking aspects in terms of infrastructure connectivity as well as protocols to access the infrastructure, and the usefulness of fog computing in support of 5G;
 \item \textbf{Management}, where management needs for the IoT-Fog-Cloud infrastructure are discussed and reviewed, including \textit{orchestration}, \textit{resource management}, \textit{services management}, \textit{energy consumption}, \textit{devices federation}, and \textit{data locality};
 \item \textbf{Applications}, where three different kinds of applications are considered, namely \textit{urban computing}, \textit{mobile applications}, and the \textit{Industrial IoT}, discussing how they can benefit from fog computing.
\end{enumerate}

% (about 1.5 page per topic, including figures,emphasizing main contributions in the literature)
%%%%% Management %%%%%
\subsection{Infrastructure}

%Cloud and Fog (Luiz Bittencourt/Omer Rana)
\subsubsection{Cloud and Fog}
    The infrastructure discussed in this paper is a composition of fog and cloud to support IoT applications, therefore constituting a three-tiered infrastructure. While IoT devices are concentrated at the edge of the network, fog devices are distributed from the IoT device's access point through the network core. The cloud is further away from the IoT sensors/actuators, requiring requests from the edge to traverse the public Internet to access cloud computing resources. As the fog infrastructure can be itself composed of different levels~\cite{openfog2017openfog}, the mid-layer of the IoT-Fog-Cloud infrastructure can offer a variety of levels of quality of service~\cite{8240187}.  Applications that have different requirements can be deployed and run on any device in this infrastructure composition, depending on their requirements. Moreover, application components can be distributed among devices at different levels of the fog depending on the application needs (e.g., latency, computing capacity, data locality).

    Cloud computing services are based on centralized data centers, where computing capacity is offered over the virtualization of computing clusters deployed in buildings specially designed to host them. Hosts in the data center are often connected through Ethernet; different topologies for this interconnect are available in the literature~\cite{Al-Fares:2008:SCD:1402946.1402967, Greenberg:2009:VSF:1594977.1592576}. Cloud users are usually not aware or concerned about details of the data center network topology, even though this can have an impact on the application's behavior. As the infrastructure management and control are provider's duties, how this infrastructure impacts application should be reflected in the Service Level Agreement (SLA) established between providers and users.
    
    Cloud data centers are large facilities deployed in a limited number of locations due to special infrastructure requirements, such as space, power, and cooling, as well as due to the need for qualified workforce and the associated management costs. On the other hand, cloud users are scattered worldwide, and consequently many users are not geographically close to cloud data centers of their preferred cloud provider. In the same way, IoT devices are scattered and may also be distant from the cloud; thus, the fog computing infrastructure can be closer to those devices to bring computing capacity with lower response time.
    
    The fog infrastructure can be organized in a hierarchy among the edge devices (IoT, mobile smart devices, etc) and the cloud data centers. The distribution (e.g., density or number of levels) of this hierarchy can vary from place to place, but the first level is expected to be located one hop away from the edge (user or device): at the access point (e.g., WiFi or cell phone antennas) or immediately above it. This would be the first (closest) offloading option for devices at the edge, providing lower latencies even though with limited computing capacity. This single level of processing can be combined with the cloud to provide the necessary computing power for applications with heterogeneous requirements~\cite{7424534}, but other fog levels may be added to enhance computing capacity closer to the edge and allow data processing/transit between devices connected to different or distant access points. The multi-tier deployment of fog nodes may depend on the use case. The number of tiers of a fog system is determined by different factors, such as the characteristics of the workload to be processed, available capacities of processing nodes, number of sensors and actuators, and latency requirements. The workload on fog nodes is generally related to the processing of latency-constrained jobs. The cloud can process heavy workloads and perform long-term storage of data. 
    
    It is common to designate the aforementioned hierarchy of computing capacity  as \textit{fog nodes}~\cite{yi2015fog}, \textit{cloudlets}~\cite{7020884} or \textit{micro data centers}~\cite{Singh:2017:RRT:3147213.3147216,7098039}.  Conceptually, the higher in the hierarchy a cloudlet is, the larger its processing/storage capacity is, since it is expected to support more devices in the tree downwards the edge. On the other hand, cloudlets that are higher in the hierarchy are also expected to present longer network delays to the edge. Therefore, the hierarchical composition of micro data centers (or cloudlets) along with the cloud provides a range of computing capacity at different geographical (and logical) distances to the IoT devices at the edge.

% Networking and fog hierarchy (device-fog: wireless; and fog-fog/fog-cloud: technologies and topologies) - (Nelson Fonseca)
% (about 1.5 page per topic, including figures,emphasizing main contributions in the literature)

\subsubsection{Networking and Fog Hierarchy}
\begin{figure}[!htbp]
\centering
        \includegraphics[width=1.0\textwidth]{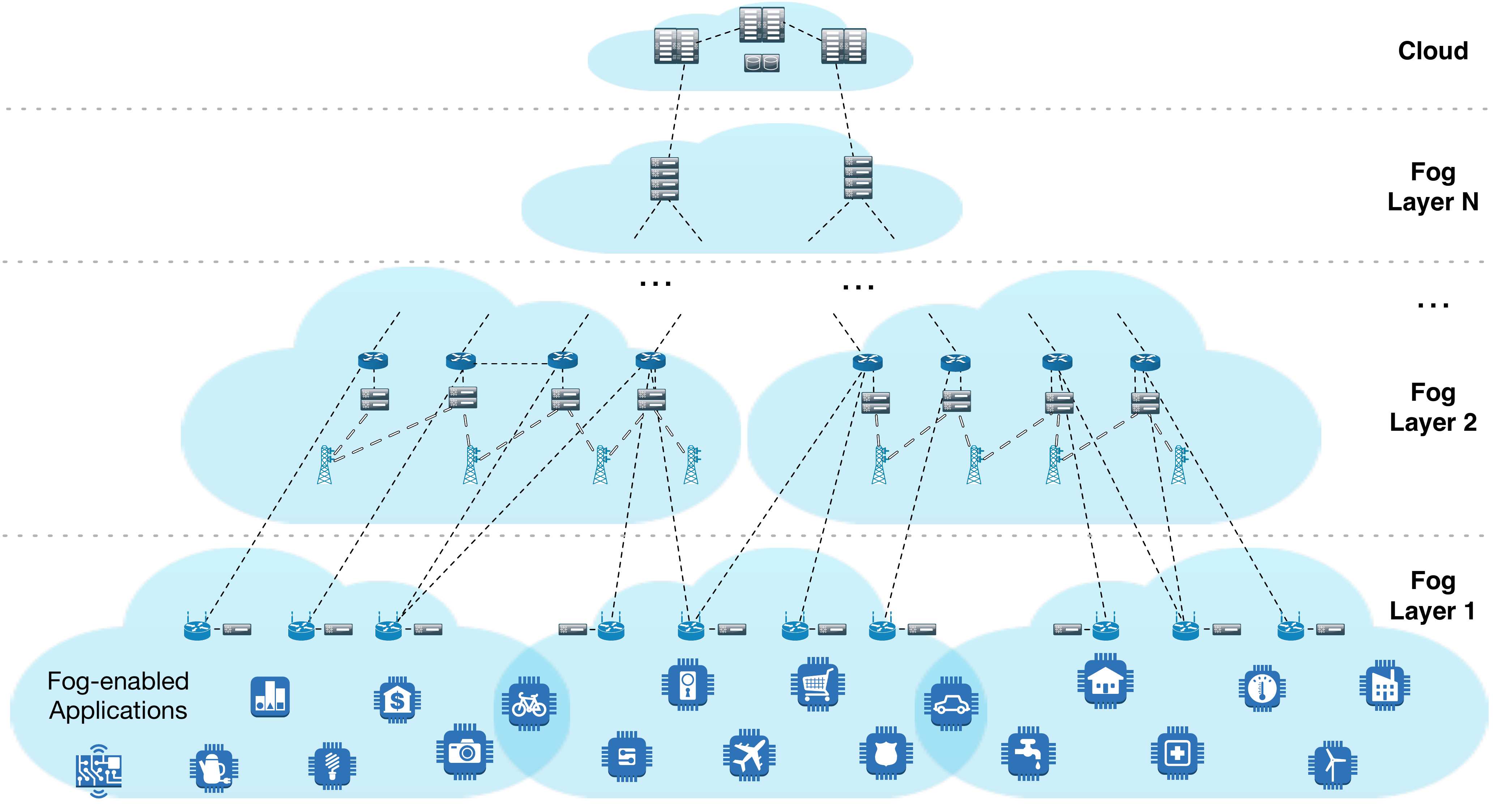}
\caption{IoT-Fog-Cloud connectivity.}\label{fig:IoTFog}
\end{figure}

%	The OpenFog Consortium has taken the lead in fog standardization and has defined a reference architecture consisting of $N$ tiers of nodes~\cite{openfog2017openfog}. Application requests come from the bottom layer, where smart end-devices perform data pre-processing and compression. The first fog tier performs a first stage analysis, passing their results to possibly intermediate fog layers for more detailed analysis, and ultimately sending their results to the Cloud~\cite{8004148}. 
The connectivity between several tiers in the fog/cloud hierarchy can be possible thanks to several network technologies, including wired and wireless ones. Figure \ref{fig:IoTFog} illustrates this fog/cloud hierarchy connectivity.
    
	The functional unit of a fog computing system, i.e., the fog node, can be either a logical or a physical entity, embedding computing, storage, and also networking capabilities. Diverse devices can act as fog nodes, including networking devices: routers, switches, wireless access points. Along with other fog nodes (e.g., video surveillance cameras and traditional servers) networking devices can also enable the processing of tasks closer to data sources, providing increased performance and security critical to health, military, or emergency applications.
        
	Data in IoT-Fog-Cloud infrastructures must traverse one or more tiers, connected by a fog network. The decision on how to connect different nodes depends on a specific technological scenario. For example, a fog node used to process raw data from sensors will typically employ wireless connections, while a fog in a factory employed to monitor manufacturing processes is likely to use wired connections. 
    
	The connection to the cloud is made by the Internet. It typically employs optical links belonging to an Internet Service Provider that will connect fog nodes to the Internet while interconnection between different fog nodes are likely to be made by Ethernet-like protocols. Moreover, the network topology connecting different fog nodes will depend on how communication channels are placed a given area, and thus different topologies and network technologies to connect fog nodes at the same or different layers can co-exist.
    
	Wireless links may typically connect IoT devices to the fog since these devices often have only wireless interfaces. Wireless connections can also be employed in fog-to-fog or fog-to-cloud interconnections depending on the available infrastructure. Cellular technologies (3G, 4G, 5G) are expected to be used in fog computing systems. For example, an architecture for fog computing, named Telcofog, has been designed to integrate fog nodes in 5G networks~\cite{8004151}. In this architecture, a fog node can be created in an edge network and accessed by end-users using 5G connections.
    
Wireless LAN (WLAN) networks are also useful in the deployment of fog nodes. Due to their limited range, they can be used in fog nodes for small buildings or campus~\cite{7987464,7511465}. Other wireless technologies for device-to-fog communication include Wireless Personal Area Networks (Bluetooth, ZigBee, Infrared) and Near Field Communication. Moreover, multi-hop wireless transmissions to route data between sensors/actuators and the fog have also been proposed in the literature~\cite{6984239}.

% Data collection and protocols (Leandro Villas)
% (about 1.5 page per topic, including figures,emphasizing main contributions in the literature)

\subsubsection{Data collection and Protocols at the Edge}

At the lowest level of an IoT network, it is expected to have hundreds or even more elements associated with each fog node.
These elements may be sensors, actuators, or control nodes, which can range from relatively dumb devices, with major processing capability constraints, to well-equipped nodes that can easily handle the full stack network protocols~\cite{Hong2013}.

Different devices may have distinct needs regarding network protocols they implement.
Because of that, it is not possible to address this issue with an one-fits-all approach. 
To make matters worse, scenarios where fog and edge computing best fit are those where data needs to be collected, analyzed, and acted upon within a few milliseconds. 
In the light of the above, the network protocols should accommodate the requirements of such communication patterns by providing the flexibility, scalability, and availability needed. 
This holds especially true for delay-sensitive applications.

Another important aspect in the IoT-Fog-Cloud hierarchy is that the communication technologies are going to be responsible to connect heterogeneous devices so that they can enable new smart services. 
The physical objects will be bridged together at the edge of the network to provide intelligent decision making capabilities by talking to each other, sharing information, and coordinating decisions without human intervention. 
It is important to notice that a considerable amount of IoT nodes are power-constrained and need to operate in noisy or lossy communication links.
Placing an excerpt of the intelligence at the fog level helps to reduce some of these device-related complexity constraints.

The most fitted network connectivity model may widely range depending on the node’s location, purpose, and specifications~\cite{openfog2017}.
Nodes may be connected through a wired network, for example, on a factory floor or other highly noisy environment.
This connection is also a good option for static nodes that require very high throughput.
One example of a communication protocol, in this case, is the Ethernet~\cite{Chiang2016}, with either copper or fiber links.
On the other hand, nodes with mobility or deployed at places without a pre-existing infrastructure can use wireless networks.
Examples of communication protocols in the latter case are Wi-Fi~\cite{Tozlu2012}, Bluetooth low energy~\cite{Chang2014}, Z-wave~\cite{Gomez2010}, IEEE 802.15.4 (WirelessHART, ZigBee, ISA100.11a, MiWi)~\cite{Lu2004}, just to name a few.

Another important advance in communications protocol technology is the IPv6 over Low-Power Wireless Area Networks~(6LoWPAN) standard~\cite{Mulligan2007}.
This technology contemplates an adaptation layer between the network and link layers to adjust the size of the packets, which are smaller in the IEEE 802.15.4 standard. 
In addition, it also implements several header compression techniques to handle resource-limited devices.
Moreover, the IPv6 allows having a much larger address space and also the capability of establishing a direct external communication link between the devices without the need for a coordinator or a gateway to translate the messages~\cite{Bouaziz2016}. 

Taking everything into consideration, having the sensors, actuators, and control nodes close together at the fog level brings several advantages. 
However, several of them may not be capable of communicating directly with a fog node.
Because of that, protocol abstraction layers are needed to logically connect all these elements as well as data collection and aggregation to consolidate the data.

The main goal of data collection and aggregation is to use a centralized approach to gather all the data in an effective manner, which provides several benefits~\cite{CUNHA2016}.
First of all, because IoT elements are close to the edge nodes, the fog can act as the first front of access control and encryption as well as provide isolation and contextual integrity.
Besides, it can also protect privacy-sensitive data~\cite{Fan2014, Jin2016}, by not allowing it to leave the edge.
In addition, sensors often generate a huge volume of raw data in burst mode, i.e., there are peaks of data transmission that fluctuate over time.
By collecting this data and aggregating the results it is possible to reduce (if not eliminate) redundancy, which leads to a decrease in network communication cost~\cite{Villas2013}.
This can also have a positive impact on the network lifespan, improve the energy consumption~\cite{Xiang2013}, prevent traffic bottlenecks, as well as enhance data accuracy by removing outliers and misreadings~\cite{Li2011}.

Another advantage of this technique is to provide data locality.
This means that it is possible to give context to the data and handle it where it makes the most sense. 
This allows making the decision process more efficient as it helps to act as soon as the raw data is converted into some meaningful context; further information can be found in section~\ref{sec:Data_management_and_locality}.

% Networking 5G (Luiz da Silva)
% (about 1.5 page per topic, including figures,emphasizing main contributions in the literature)

\subsubsection{An Infrastructure to Support 5G}

Current developments in 5G are characterized by the orchestration of network resources to meet a wide range of services, broadly classified into three categories: enhanced mobile broadband (eMBB), ultra-reliable and low-latency communications (URLL), and massive machine-type communications (mMTC). While a fog/cloud computing hierarchy plays a role in all three, IoT services are generally considered under the mMTC heading. The idea is that a 5G operator will be able to slice the network, provisioning each slice to meet the very diverse needs of the three types of service, in terms of latency, reliability, throughput, scalability, and mobility support. A slice, in this context, can be thought of as a virtual network, whose resources are provisioned for a particular service or class of service and isolated from other slices that share the same physical infrastructure.
They are expected to be one of the key resources in 5G networks~\cite{Zhang2017}, by providing a holistic end-to-end virtual network for a given user, so-called tenants.
This means that the physical mobile network will have its resources partitioned and customized according to the system needs~\cite{Samdanis2016}.
%The combination of two technologies are envisioned to offer support to the network slices, namely Network Functions Virtualization~(NFV) and Software-defined Networking~(SDN)~\cite{Yousaf2017}.
%The former allows instantiating on-demand specific network functions and services, which enables load balance, failure recovery, and better hardware utilization.
%The latter decouples the control and data planes, making it possible to abstract the underlying network infrastructure.
The outcome is a more flexible, reliable, scalable, and secure network.
Using these technologies, in many situations, the networks will be able to reconfigure slices within seconds to quickly respond to local demands, such as an unexpected gathering of people or to prioritize emergency systems. 
On the other hand, it is also possible to program a long-term lease, for example, to an electrical utility company to accommodate its smart grid components such as meters, sensors, controllers, and other IoT devices. 
A short-term lease is also feasible, for example, when a public venue or a concert promoter wants to have a dedicated slice for a weekend-long festival and optimize it for streaming high-quality video and music data. 
%Therefore, to the provisioning of resources from the things to the cloud passes through the fog.

While it is possible to envision a network slice to support a particular IoT service throughout the fog hierarchy, it is the ability to combine and process vast amounts of data in the cloud, originating from multiple IoT services through the fog, and to apply machine learning techniques to those data, that opens up potentially revolutionary progress in a myriad of fields, from healthcare to precision agriculture.

Edge devices in IoT are often resource-constrained in energy storage and processing capabilities. The combination of cloud and fog computing provides some relief to these limitations, meeting requirements of 5G such as geo-distributed real-time processing and runtime adaptability~\cite{7495388}. As the fog computing paradigm emerges, the development of 5G architectures with fog support started to appear. For example, fog radio access network (F-RAN) was proposed to combine communications and computing operations for 5G~\cite{7901475}. 

One of the intrinsic challenges then becomes how to achieve the low latency requirements of real-time IoT services. Recent work suggests the adoption of a hybrid fog/cloud solution in order to support latency-sensitive IoT services~\cite{yannuzzi2014key,7548876}. The impact of the lack of processing power on the delay performance of an IoT service that relies on voice and gesture commands from the end-user to control a set of lights in a smart home or office has already been demonstrated ~\cite{8116434}.

%%%%% Management %%%%%
\subsection{IoT and Fog Management}
% Resource allocation and optimization (Rizos Sakellariou)
% (about 1.5 page per topic, including figures,emphasizing main contributions in the literature)

\subsubsection{Resource Allocation and Optimization}
\label{sec:allocation}
   Resource allocation has been a challenging problem in distributed systems and, as novel infrastructures appear, new variables must be considered. Data about application requirements and infrastructure characteristics are taken as input to optimize an objective function in order to map applications to the resources available in the infrastructure. A scheduler is the entity responsible for running an optimization model that takes those data as input and generates an application schedule into resources as an output trying to maximize or minimize a single objective or a set of (conflicting or not) objectives.
   
   As the scheduling problem is NP-Complete in general~\cite{pinedo2016scheduling}, many different techniques to model and solve it have been proposed in the literature~\cite{kan2012machine}. Notwithstanding, with the emergence of IoT, a plethora of devices and applications have been suggested, bringing the heterogeneity in both application requirements and infrastructure characteristics to unprecedented levels. The scheduling literature in distributed systems shows that scheduling models and optimization techniques are sensitive to applications and infrastructure characteristics~\cite{1558639}. Therefore, when heterogeneity becomes the norm, as it is in IoT, schedulers should be able to adapt to different scenarios, or multiple schedulers should coexist to handle optimization models with different characteristics and/or objectives.
   
   Currently, IoT applications commonly rely on cloud computing to process and store data. Resource allocation in cloud computing can be seen from two different perspectives: (i) allocation of resources within the cloud provider's data center; and (ii) allocation of applications to services offered by cloud providers. Perspective (i) is often referred to as a \textit{VM placement} problem, which is an optimization problem that aims at distributing virtual machines in a data center~\cite{meng2010improving,Pietri:2016:MVM:2988524.2983575}. Objective functions common for the VM placement problem are to maximize the utilization of the data center, minimize network traffic, and minimize energy consumption~\cite{li2013energy,meng2010improving,Pietri:2016:MVM:2988524.2983575}. Perspective (ii), often referred to as \textit{application scheduling}, is concerned with matching application requirements with services by taking into account application requirements and service level agreements from cloud providers. Common objectives in application scheduling are minimization of the execution time~\cite{pandey2010particle} and minimization of costs~\cite{6295710}, as cloud computing services charge on a pay-per-use basis.
   
   Fog computing is expected to fulfill requirements that cloud data centers are not able to, but yet rely on the cloud when such requirements do not exist or are not mandatory. In this sense, with fog models and architectures under discussion in the literature, research on schedulers that take fog infrastructures as input has been carried out in the last few years~\cite{zeng2016joint,7912261}. One of the main questions that arise is how to distribute heterogeneous data, jobs, and services throughout the fog/cloud hierarchy in a way that application requirements are met and infrastructure utilization is efficient, avoiding bottlenecks as the resources closer to the edge in the fog infrastructure are constrained. In the IoT landscape, requirements such as latency can play an important role in the scheduler decision making, thus being a determinant of where applications should run~\cite{Bonomi:2012:FCR:2342509.2342513}. Moreover, the combination of heterogeneous applications with mobility has also been the focus of attention lately~\cite{7912261}, as dynamic demands on edge nodes call for dynamic approaches to redistribute the load in the fog/cloud hierarchy.

% Services (Microservices / Serverless Computing/ Osmotic Computing / Social Computing) - (Omer Rana)
% (about 1.5 page per topic, including figures,emphasizing main contributions in the literature)

\subsubsection{Serverless Computing}

The new computing systems that are being proposed have to be able to handle a myriad of heterogeneous devices, which can often have different computational capabilities. Furthermore, there is a varying network, data communications bandwidth, and latency available closer to a data center compared to the network edge. Understanding which services should execute on a cloud data center and which on the edge devices remains a challenge. 

The serverless perspective focuses on the provision of computational {\it functions}, with limited resource requirements, that can be deployed closer to user devices such as AWS Lambda, Google Cloud functions, and Microsoft Azure functions.
In these systems, the functions are triggered based on user-defined events.
In the AWS case, it can use of a number of other AWS services, such as DynamoDB and S3, just to cite a couple. AWS lambda involves hosting such functions using Amazon CloudFront, which means that it will be used by a data center in close proximity to the data source. A key benefit in adopting these approaches involves rapid deployment of lambda functions as well as the limited resource needs for executing them. This approach has received significant traction for real-time data streams processing. By doing that, a data feed can be shared and distributed for processing across multiple functions through AWS Kinesis. The serverless approach also modifies the traditional public cloud systems approach, from batch-oriented processing to a close to real-time processing of data. 

Serverless computing is expected to grow across the IoT-Fog-Edge-Cloud systems as an extension of current cloud-based implementations to support IoT applications. 
In this direction, EdgeScale~\cite{7774691} aims to implement a serverless computation model that enables scalable and persistent storage services to be scattered through a data center hierarchy, thus compatible with the fog computing architecture. The focus of EdgeScale is to enable automatic application state movement through the hierarchy considering that applications can run at different levels depending on user needs and current network status.

In the same direction, the Osmotic Computing (OC)~\cite{7802525} focus has been on developing microservices which can be migrated from cloud data centers through the fog/edge driven by performance and/or security constraints. The OC model specifically focuses on creating {\it migratable} hosting environments for microservices that can be moved dynamically, taking account of device specific characteristics of edge resources. It suggests the use of frameworks such as JaJa, Fabric8, and PXE for hosting container-based deployment of microservices to support latency-sensitive applications. It also suggests the use of a serverless style of processing, however, the processing functions are deployed using migratable containers. Many serverless capabilities currently available also rely on vendor specific solutions (e.g., AWS lambda), which are hard to generalize across vendors. On the other hand, the OC perspective is vendor-neutral, relying on identifying a function hosting approach that can be shared between different vendors.

% Energy consumption (Rizos Sakellariou)
% (about 1.5 page per topic, including figures,emphasizing main contributions in the literature)

\subsubsection{Energy Consumption}

There are two key aspects to be considered when discussing energy consumption in relation to IoT-Fog-Cloud computing. On one hand there is a plethora of studies that argue for the benefits that IoT can bring in reducing energy consumption in various settings~\cite{7061425,7805265,5940920,LEE2015431}.
Some of these benefits in terms of energy are also discussed later in the section on Urban Computing.
On the other hand, however, there are strong indications that, on their own, IoT-Fog-Cloud technologies may lead to additional pressures in energy consumption. Clearly, a full evaluation would need to weigh costs and benefits: a small increase in IoT or cloud energy costs may be offset by significantly higher savings in the domains where these technologies are applied. To the best of our knowledge there has been no such holistic assessment in the literature but it may take some time until comprehensive studies in this respect may be produced.

Pressures in energy consumption can be addressed at primarily three levels: (i) hardware and infrastructures in general; (ii) systems software; (iii) data management. In terms of hardware and infrastructures it is useful to distinguish between work to produce energy-efficient chips or devices and work at a level that considers infrastructures built with multiple such chips or devices. The former consists of various energy-efficient architectural features that may include techniques for voltage scaling or cooling \cite{6291714,Dabbelt:2016:VPE:2934495.2934497,7284406,7927716,Luo:2018:NDT:3195970.3196080}. The latter includes the large body of work aiming at minimizing the energy cost of running large-scale infrastructures, data centers in particular~\cite{Urgaonkar:2011:OPC:1993744.1993766,BELOGLAZOV201147,7373667,7279063}.

A number of researchers have focused on systems software issues, particularly resource management. As IoT-Fog-Cloud platforms typically consist of multiple and heterogeneous resources, efficient use of such resources can make a significant impact on energy. In fact, energy has often become a key consideration in a variety of mapping and resource allocation techniques \cite{BELOGLAZOV2012755,Hameed2016,Pietri:2016:MVM:2988524.2983575}, with some work focusing on resource management for specific applications or services~\cite{7217791,6686006,BAKER201796,7448886}. Somewhat orthogonally, the energy consumption impact of different programming languages has been assessed in \cite{Georgiou:2018:YPL:3196398.3196414}. It is expected that this line of research will intensify in the broader IoT-Fog-Cloud context.

Significant energy savings could be obtained by carefully managing the large size of data that IoT-Fog-Cloud applications may potentially generate. Different approaches may include: (i) algorithms for energy-aware data transfer~\cite{Alan:2015:EDT:2807591.2807628}; (ii) algorithms that trade computation with communication, as in \cite{Pietri:2018:SDS:3221269.3221298}, possibly using strict energy consumption objectives to balance this trade-off; or (iii) algorithms that limit the amount of data that could potentially be produced and transmitted through network links using some sort of satisfaction criterion~\cite{lambert2018}, which may also include energy consumption thresholds that should not be exceeded. Overall, handling data in an economical manner and avoiding costly communication as much as possible in the IoT-Fog-Cloud continuum (see next section) may lead to significant energy savings.

%Data locality (Roger Immich)
% (about 1.5 page per topic, including figures,emphasizing main contributions in the literature)

\subsubsection{Data Management and Locality \label{sec:Data_management_and_locality}}

In recent years, there has been a considerable increase in the creation and consumption of data, which has reached unprecedented rates. As a result, data management and locality have been gaining significant attention lately.  
These concepts have been researched in different contexts in cluster, parallel and distributed computing in the past. However, they had not been adopted to geo-distributed data centers such as Cloud-Fog-Edge computing, until recently~\cite{Xu2013,Hung2015,Heintz2016}.
In principle, the two concepts refer to the capability of organizing and maintaining data-related processes, which include acquiring, processing, distributing, storing, protecting, and validating information. 

Data management involves the design and deployment of policies, architectures, and procedures allowing the accurate management of the full data lifecycle. 
It broadly relates to two generic strategies, the placement strategy and the access strategy~\cite{Sakr2011}.
The first one defines where and how the data should be distributed.
This includes defining how many copies should be made and what the best nodes are to store these copies. 
The second one prescribes how the read and write operations are going to be handled by the system.
This strategy has to take into account consistency among the distributed copies and how each copy is going to be accessed through the network. 

Data locality is related to the capability of moving the computation close to where the actual data is being created or acquired, rather than transferring large amounts of data to a centralized computational resource~\cite{Yang2017}.
This concept goes against pushing all the indiscriminate raw data directly to the cloud.
This is based on the fact that it is cheaper, in terms of network resources, and more efficient to move and execute a computational application near to the data it operates on.
This is one of the main premises of the fog computing: to have a decentralized system with resources close to the end-users~\cite{7867735}.
This is especially true if the data to be analyzed is considerably large.
In doing that, it is possible to reduce the core network congestion and decrease the latency, as well as to improve the overall throughput of the system~\cite{Yi2015}.

Nowadays, data-intensive applications are increasingly relying on geographically distributed resources to store and process information. 
Several frameworks have been proposed to make an efficient use of large computational clusters providing massive data processing capabilities.
The MapReduce framework~\cite{Dean2008}, for example, is able to schedule jobs taking into consideration the data locality.
This means that the jobs are divided into several tasks and then dispatched to the node which has the data to be processed.
Doing this reduces the network overhead by avoiding unnecessary data movement and improves the individual job feedback time within a cluster, as well as decreases the latency~\cite{Greenberg2008,Vulimiri2015}. 

In fog computing, data locality issues have been addressed more recently, indicating that data storage at the edges in a fog can improve response time and reduce network traffic~\cite{Confais2017}. 
Another positive aspect of data locality in fog computing is the enhancement of security- and privacy-related issues~\cite{Bellavista2017}.
By operating locally, it is possible to have an accurate knowledge of the gateways and easily implement authentication and authorization features.
At the same time, because the data and the processing nodes are close to each other, the information does not need to be moved around, thus facilitating privacy~\cite{Vaquero:2014:FYW:2677046.2677052}.

% Orchestration (Marilia Curado)
% (about 1.5 page per topic, including figures,emphasizing main contributions in the literature)

\subsubsection{Orchestration in Fog for IoT}
%The introduction of a fog layer to support the Internet of Things has the potential to bring several advantages such as low latency, mobility support and increased resilience. Going one step further from the cloud-based IoT, where resources are centralized in large data centres requiring extensive and expensive communications with IoT devices, the fog follows a distributed approach, splitting resources among multiple smaller entities which are much closer to the edge devices.  
As mentioned earlier, the fog can be organized in multiple layers and spread throughout different entities creating a highly dynamic, large-scale, heterogeneous and complex scenario. Within this ecosystem new challenges arise in terms of dynamic resource management and orchestration functions. These challenges are being addressed in different ways in the literature, as described in this section. 

The main objective of orchestration functions comprises the dynamic management of resources considering applications requirements and the related workloads characteristics, which, in many cases include a transient operational behaviour. Fog resources are manifold involving basic sensors, CPU and memory components, virtual machines and virtual network functions, as well as network and applications services and micro-services. The role of fog orchestration is thus to guarantee the proper functioning of all these resources while guaranteeing security and an adequate application performance level. 

To achieve its objectives, a fog orchestrator must perform the following functions~\cite{Velasquez2018}:
\begin{itemize}
\item Scheduling and placement -- The main role of scheduling and placement is to decide which applications should be executed where and when~\cite{Velasquez2017, Skarlat2017}. To fulfill this goal a range of information must be considered, such as applications requirements, resource availability and mobility patterns, as discussed in Sections \ref{sec:allocation} and \ref{sec:mobility}. In the edge, the orchestrator should also make the scheduler aware of the need for migration of tasks and data.
\item Discovery and allocation -- To support scheduling, it is of utmost importance that the orchestrator has updated information about the resources and devices in the fog~\cite{7912261}. In addition, resource allocation must be performed according to optimization criteria that suit applications requirements. Multiple trade-offs arise in this context due to the complexity of the functions involved; the updated information provided by the orchestrator drives the scheduling algorithm towards the optimization function.
\end{itemize}

Several fog orchestrators have been proposed in the literature with different objectives, such as reducing latency, improving resilience, ensuring security and privacy, among many others, as summarized next.

GA-Par (Genetic Algorithm Parallel) has been conceived to manage IoT application composition using a genetic algorithm~\cite{7867735}. The main application requirements considered were security and network QoS.  Although this work provides insights towards orchestration, it presents scalability issues highlighting the challenges associated with dynamic adaptations within fog-based IoT orchestration. 

The ECHO middleware platform provides orchestration capabilities for data flow composition over distributed resources, including edge, fog and cloud~\cite{Ravindra2017}. In addition, ECHO also supports task migration, which is fundamental to adapt to the dynamicity of these systems. However, the proposed approach has a centralized nature which limits its applicability in large-scale systems as well as its resilience. 

Overcoming the well-known limitations of centralized approaches, CF-Cloud Orch (Cloud Fog Orchestration) proposes a distributed solution for cloud orchestration using container-based fog nodes and a SDN management system~\cite{Kim2018}. The main management functions supported include security, scheduling and load balancing. Although the proposed architecture aims to be scalable, the paper does not present results to support such an objective.

A new trend to address the challenges of fog environments combines orchestration with choreography~\cite{8114500}. This hybrid approach resorts to service orchestration for resource management between the fog and the cloud and service choreography for resource management between the IoT devices and the fog. On one hand, with the orchestration approach it is possible to have a global view of the environment and to efficiently use and manage the fog/cloud resources. On the other hand, a finer view of the choreography supports more efficient localized decisions.

In addition to the related work presented above, there are several initiatives for the standardization of management and orchestration functions in fog environments. The European Telecommunications Standards Institute (ETSI) has ongoing efforts in the Mobile Edge Computing (MEC~\footnote{https://www.etsi.org/technologies-clusters/technologies/multi-access-edge-computing}), and NFV Management and Orchestration (MANO~\footnote{https://www.etsi.org/technologies-clusters/technologies/nfv/open-source-mano})  technical committees. In addition, the OpenFog Consortium has designed an architecture for fog computing which has been adopted by IEEE \cite{openfog2017openfog}.  Two fog orchestration architectures that closely follow the activities of these standardization bodies have been proposed within the context of 5G networks~\cite{7946419,santos2017fog}. While the first paper is still at architecture level, the second paper has been assessed in a smart cities scenario comprising autonomous fog node management, data analysis and decision making functionalities. 

%This section has provided insights on the recent developments in the area of fog orchestration for the Internet of Things, highlighting its major challenges and open issues being investigated by the research community and standardization bodies. 

%Federation, authentication, identity, security (Craig Lee)
% (about 1.5 page per topic, including figures,emphasizing main contributions in the literature)

%\subsubsection{Devices federation, authentication and identity management\textbf{(Craig Lee)}}
\subsubsection{Applying Federation Concepts in Fog and IoT Environments}

The Internet of Things and the notion of fog computing is the natural
evolution of interconnectedness that is part of this integrated fabric
that is touching every aspect of academia, industry, government and
culture.
This interconnectedness is being driven by the need to {\em collaborate},
i.e., to share data and resources.
In more established computing environments, this has led to the
development of networked communications, the World Wide Web, and all
manner of social media.

As pervasive as the technologies are, there has still been a need to
manage shared resources in a more intelligent, comprehensive manner
that is less {\em ad hoc}.
This has given rise to the concept of {\em federation}.
In short, a federation is a security and collaboration context wherein
participants from different organizations and administrative domains
can jointly define, agree upon, and enforce joint resource discovery
and access policies \cite{Lee2016}.
It is clear that fog computing environments and IoT devices will
ultimately need some type of federation to manage how different sets
of data producers and data consumers can collaborate and share data.

Just as one example, consider a smart electric vehicle on a road trip.
Will the vehicle owner authorize a local power company to monitor the
geographic location and current charge of the vehicle to (a) direct
the vehicle to an appropriate charging station when it needs to
recharge, and (b) ensure that sufficient electrical power is available
at the charging station at a given cost?
This one example includes multiple aspects of policy, authorization, trust
and also mobility.
The vehicle owner may have authorized their ``home'' power company to
monitor charge, but another entity may have to be authorized to monitor
location and share this information with the power company.
The home power company may have to delegate the responsibility for
re-charging to the ``local'' power company wherever the vehicle
actually happens to be.
This local power company may have to report the charging cost to the
home company such that the vehicle owner can be billed appropriately.
All of this needs to be securely managed such that the vehicle owner
can manage their privacy as desired, and that electrical power
production and consumption is efficiently managed and fairly paid for.
In general, federations can provide the virtual context wherein such
policy agreements can be made and enforced.

The fundamental importance of federation was clearly recognized and
articulated in the NIST Cloud Computing Technology Roadmap
\cite{NIST:CCTR1} as {\em Requirement 5: Frameworks to Support
  Federated Community Clouds}.
The concept of using federation to manage collaboration, however, is
not cloud-specific.
Over the last fifteen years or so, a tremendous amount of work has
been done to support different aspects of federations, and for
different specific use cases.
Systems such as InCommon \cite{InCommon} and eduGAIN \cite{eduGAIN}
were developed to enable metadata about Identity Providers and Service
Providers to be exchanged among participating organizations.
The Interoperable Global Trust Federation \cite{IGTF} was formed to develop
trust criteria and enable IdPs to be trusted among participating
organizations.
Globus Auth \cite{GlobusAuth2016} was developed whereby IdPs and SPs
can be managed with more structure, and a user's Globus Auth
credentials can be delegated to a third-party SP to act on behalf of
the user.
There are many more examples, but a more complete review is outside
the scope of this paper.

While certainly important and widely used, these systems were also
developed in an {\em ad hoc} manner.
Different deployment and governance mechanisms are static and
``baked-in'' to their designs and operation.
While standards, such as SAML, OpenID, OAuth, and OpenID Connect, are
used for their specific purposes, the federation capabilities themselves
are not built to any standards.

To address this issue, NIST and the IEEE formed a Joint Federated
Cloud Working Group \cite{WG_KO}.
The NIST goal is to clearly define the {\em federation design space}
in a Reference Architecture that identifies the major {\em actors}
involved in general federations and how they can and must interact.
(See \cite{NIST_Fed_Ref_Arch} for more details.)
Areas of desirable, {\em federation-specific} standards would be
taken through the international process by the IEEE.

Clearly the vast majority of work done in federation has been done in
established computing environments.
The question at hand is {\em how well can these unifying, federation
  concepts being developed by NIST be applied in fog computing
  environments and IoT devices?}
IoT devices will be numerous, highly distributed, and typically
operated in resource-constrained environment.
The term {\em fog computing} was coined to denote computing resources
that are ``closer to the ground''.
As such, fog computing is commonly described as having a three-tier
model of cloud resources, fog nodes, and IoT edge devices
\cite{Yi2015, Stoj2015}.

All of these computing resources and devices will exist in inherently
distributed environments and could have different owners.
Owners will typically want to manage how their data is produced, how
it is collected, how it is consumed, and by whom.
As such, these resources, devices and data will be managed in
different {\em administrative domains}.
However, as much as any other organization in a connected world, IoT
device and data owners may have strong motivations to collaborate and
share data with other organizations.

The core goal of the NIST/IEEE Joint WG is to clearly define how the
sharing of resources and data can be done across administrative
domains in a general, standardizable way.
The key concept is the ability to establish and manage {\em
  virtualized administrative domains} that may span multiple
organizational administrative domains.
Such virtualized domains are {\em federations}.

The draft NIST Cloud Federation Reference Architecture
\cite{NIST_Fed_Ref_Arch} is based on the notion of a {\em Federation
  Manager (FM)}.
This FM is capable of a specific set of fundamental capabilities to
support federations.
However, FMs can be deployed singly or large groups in a wide range of
deployment and governance models.
The choice of deployment and governance model can be driven by the
requirements of the application domain.
These models are described by the following properties:

\begin{itemize}
\setlength\topsep{0pt}
\setlength\itemsep{0pt}
\setlength\parskip{0pt}
\item Deployment/Scale Properties

\begin{itemize}
\setlength\topsep{0pt}
\setlength\itemsep{0pt}
\setlength\parskip{0pt}
\item Internal vs. External FMs.
%  Using an external federation provider will certainly be a common use
%  case, but using internal FMs avoids the additional trust
%  relationships necessary for external FMs.

\item Centralized vs. Distributed FMs.
%  Having one centralized FM is certainly simpler than having a large
%  number of FMs that effectively operate as a large distributed FM.

\item Simple vs. Large/Arbitrary Communication Topologies.
%  Simple, pair-wise, or point-to-point federation topologies that are
%  manually managed are certainly simpler than large, essentially
%  arbitrary topologies that may be built-up from many disparate sites
%  that wish to join a federation.

\item Homogeneous vs. Heterogeneous Deployments.
%  Deployments can be significantly simpler if “the same code is
%  deployed everywhere”, but larger deployments will necessarily
%  involve heterogeneous FM implementations.
\end{itemize}

\item Governance Properties

\begin{itemize}
\setlength\topsep{0pt}
\setlength\itemsep{0pt}
\setlength\parskip{0pt}
\item Implicit vs. Explicit Trust Relationships.
%  How to determine if Federation Managers can trust each other, one
%  way or another.

\item Vetting/On-Boarding New FMs.
%  How to determine if a new Federation Manager can be added to an
%  existing trusted set of FMs

\item Federated Identity.
%  How to establish a commonly understood identity when credentials may
%  come from different administrative domains.

\item Roles/Attributes.
%  All federation must have some set of roles or attributes whose
%  semantics is commonly known.

\item Federated Resource Management:
\begin{itemize}
\setlength\topsep{0pt}
\setlength\itemsep{0pt}
\setlength\parskip{0pt}
\item Discovery.
%  The must be some mechanism whereby the resources being made
%  available in a specific federation can be discovered by the
%  federation members.

\item Discovery Policies.
%  Federation members should only be able to discover resources they
%  are authorized to use, as determined by discovery policies based on
%  the roles/attributes defined within a federation.

\item Access Policies.
%  Similarly resources should have access policies defeind over the
%  roles/attributes defined within a federation and granted to specific
%  federation members.
\end{itemize}

\item New Federation Member Vetting/On-Boarding.
%  A candidate federation member must go through a vetting process
%  against some set of qualifications for membership.

\item Accounting/Auditing.
%  As federations become larger, there will need to be some way of
%  managing and verifying the exchange of value.

\item Federation Discovery.
%  While some federations may not want to be discoverable the potential
%  members, in general, many federations will wish to be discoverable
%  according to some discovery policies.

\end{itemize}
\end{itemize}

It should be clear that federation deployments can range from very
simple, single FM deployments, to global-scale, hightly distributed
deployments.
Likewise, the governance properties may be greatly simplified.
In some cases, the need may just not exist.
For example, simple federation deployment may not need to have a
general federation discovery mechanism.
When federations become more common and widely deployed, however,
having such a discovery service would be very useful.
Simple, small-scale federations may also have no need for accounting
and auditing.
Federations could also operate using out-of-band methods for agreeing
on identity credentials, roles/attributes, resource discovery, and
resource access policies.
As federations become larger and involve more amounts of significant
resources, then more formal methods for addressing these governance
requirements will need to be in place.

These deployment and governance properties will take on additional
dimensions when considering fog and IoT environments.
In a three-tier model, IoT devices will typically be designed for
minimal power requirements.
As such, they will typically not have any extra capacity for hosting
the federation properties itemized above.
Fog nodes could host these functions, depending on their actual
capacity.
Fog nodes could possibly run their own, internal FM, but could also
simply act as a client to an external FM that is hosted in a fully
functional data center.
If a fog node is only managing a small number of IoT device types,
then it may be possible that roles/attributes and resource
discovery/access policies can be statically defined out-of-band.
In any case, scale will certainly be an issue.
A fog node will have a finite capacity that will define how many IoT
devices it can manage and how it can manage the sharing of data with
external data consumers.

%==================================================

\subsubsection{Trust Models to Support Federation in Fog and IoT Environments}

The fundamental requirement for all of these deployment and governance
properties, however, is having a {\em trust relationship} among the
administrative domains, i.e., federations.
Once trust is established, the governance mechanisms can be put in
place.
Hence, how federations can be managed in fog and IoT environments will
depend on how trust can be managed.
This leads us to our central question: {\em What are the possible ways
  of managing trust, and how well would they work in fog and IoT
  environments?}
We review here a number of ways of looking at this issue.

{\bf Implicit Trust.}
For completeness we note that informal federations can be formed using {\em implicit trust}.
That is to say, if two or more organizations already have a working
relationship, an implicit trust relationship already exists.
Based on such implicit trust, these organizations can easily configure
their Federation Managers to interact with each other, and support
useful collaborations.
An example of this is the OpenStack Keystone OS-FEDERATION API
extensions \cite{OS_Keystone_Fed_API}.
The administrator of a Keystone deployment can configure their
Keystone to trust another Keystone either as an IdP or SP.

{\bf Trusted Identity Providers.}
More formal federations rely on {\em trusted Identity Providers}.
The Interoperable Global Trust Federation (IGTF) was created to
facilitate this reliance on a global scale \cite{IGTF}.
IGTF defines a minimum set of requirements and recommendations for the
operation of Public Key Infrastructure (PKI) Certificate Authorities.
IGTF maintains a set of {\em authentication profiles} concerning
things like attribute assertions and attribute release.
Once an IdP demonstrates that it complies to IGTF requirements, then
other organizations will trust the certificates signed by that IdP.

Set of trusted IdPs are also managed by systems like InCommon and
CILogon.
InCommon \cite{InCommon} maintains a metadata catalog of IdPs and SPs
that were vetted when added to the catalog.
Globus Auth \cite{GlobusAuth2016} can rely on InCommon and other IdPs
to manage {\em linked identities}.
CILogon \cite{CILogon} (which lives within the InCommon ecosystem)
can rely on 80+ different IdPs when providing federated PKI
certificates to users based on their home identity.
Organizations such as XSEDE, OSG, and LIGO all rely on CILogon
certificates to make data and services available to their users, based
on CILogon certificate attributes.

The G\'EANT Trusted Certificate Service (TCS) performs a similar
function, but within the realm of G\'EANT services \cite{GEANT_TCS}.
TCS runs the Trusted Academic CA Repository (TACAR), which hosts the
PKI trust anchors needed for G\'EANT services such as eduroam, eduGAIN
and perfSONAR.
We note that IGTF has accredited most of the CA root certificates
hosted by TACAR.

%--------------------------------------------------

{\bf Blockchain.}
A Blockchain is a {\em replicated ledger} that uses cryptographic
techniques and consensus algorithms to build trustworthy systems in an
otherwise trustless world \cite{Zheng2018BlockchainCA}.
A blockchain is simply, as the name implies, a chain of blocks, or
data structures.
Each block contains a cryptographic link to the previous block.
Each block consists of a header and content.
The header contains a link to the previous block, a time stamp, and a
Merkle hash value.
The Merkle hash value is cryptographically linked to the entire
contents of the tree.

The blockchain ``magic'' happens in how new blocks are added to the
chains, i.e., how a {\em consensus algorithm} is used to establish
agreement among participants for adding a new block.
Proof of Work is the most common consensus mechanism used in
blockchain implementations.
This proof of work in early blockchain implementations required
blockchain miners to experimentally determine which cryptographic
nonce makes the hash of its current header fit the current target.
Proof of Stake gives advantage to those miners that have a larger
stake in the blockchain ecosystem.
The Practical Byzantine Fault Tolerance (PBFT) algorithm provides a
lower latency mechanism where arriving messages are signed, and if
enough identical messages are received, then consensus is achieved.
Other consensus algorithms are possible that all have their pros and
cons with regards to cost, throughput and scalability.

%--------------------------------------------------

{\bf Named Data Networks.}
%  Material from VO-NDN paper:
Named Data Networks (NDNs) or Information-Centric Networks (ICNs)
represents not only a different way of managing communication, but
also trust.
Rather than managing communication based on fixed machine addresses,
i.e., IP addresses, all communication is based on a {\em hierarchical
  name space}.
The NDN concept was, in fact, developed to address IoT requirements
with billions of diverse devices with relatively small data messages,
yet in massive volumes in a resource-constrained environments
\cite{Amadeo2015}.

In NDNs, an {\em Interest packet} are issued for a given name path
that are routed among {\em NDN Forwarders}.
When the named data is found, a Data packet is returned.
A key design principle of NDNs is that all data packets are
encrypted.
Rather than relying on a secure channel or session, NDNs
encrypt all packets and manage the key distribution such that only
entitled users can decrypt the appropriate data packets.
Attribute-based encryption approaches can be used whereby the keys
used to encrypt and decrypt a ciphertext are derived from the data
access policy \cite{Sahai2005,Bethencourt2007}.

A method of automating the decision about which keys can sign which
data and how signature verification can be done is discussed in
\cite{Yu2015b}.
This process is facilitated by the use of {\em trust schemas} which
are defined by a set of {\em trust rules}.
Within the name space, each trust rule defines a relationship between
the name of the data and its signing keys.
This could be through a shared prefix, shared suffix, or common name
elements at specific positions.
We note that trust schemas must nonetheless reference one or more {\em
  trust anchors} that are pre-authenticated using out-of-band
mechanisms.

\vspace{3mm}

Aside from federations based on implicit trust, we note that all of
the other trust mechanisms rely on cryptographic methods of some
sort.
While implicit trust relationships may be feasible for some
applications, in general, stronger trust and security mechanisms will
have to be used.
For completeness, we can consider {\em reputation systems}
\cite{Josang07}, but these must also rely on establishing identity of
one type or another.
It is claimed that reputations can be established even for anonymous
parties as long as they can be recognized from one interaction to
the next.
Nonetheless, reputation systems only provide was can be called {\em
  soft security} which will be insufficient for most fog and IoT domains.

Hence, in general, most federations will employ cryptographic methods
to establish identity and trust.
This implies that applying federation techniques to managing fog and IoT
devices will require a sufficient degree of computing resources to
support cryptographic or consensus operations.
It is reasonable to assume that fog nodes will be capable of this,
while IoT devices themselves will not be.
This implies that security between the fog nodes and the IoT devices
must be secured at the hardware communication level.
On the fog node, however, a number of different trust mechanisms could
be employed that, in turn, enable a number of federation governance
models.
This does, however, raise some fundamental questions:
{\em How lightweight can cryptographic methods be made such that federation
mechanisms can be moved further ``down the stack'' to less powerful
and capable fog nodes, and closer to the IoT devices themselves?} 
Likewise, since different deployment and governance models can be
applied to different federation instances, {\em can more lightweight
  models be devised that minimize the need for cryptographic
  operations?}
These are outstanding issues for using federations to manage fog and IoT environments.

%%%%% Application %%%%%
\subsection{Applications}

%Urban computing (Leandro Villas)
% (about 1.5 page per topic, including figures,emphasizing main contributions in the literature)

\subsubsection{Urban Computing}
Urban computing is the process of acquiring, integrating and analyzing a large volume of heterogeneous data produced by various sources in urban spaces; for instance, sensors, vehicles and human beings to tackle various problems that cities face such as air pollution, public safety, urban mobility, lack of water and increased energy consumption. Thus, one of the main objectives of that area is to help improve the quality of life of people living in large urban environments~\cite{zheng2014urban}. In this context, fog computing is expected to help in data processing and storage for knowledge extraction to solve long- and short-term urban-related problems. 

According to the United Nations, nowadays, 55\% of the world's population live in urban areas, a proportion that is expected to increase to 68\% by 2050. Consequently, there is an enormous pressure on providing the proper infrastructure to large cities, such as transport, housing, water, and energy. To understand and partly tackle these issues, urban computing combines various data sources such as those coming from the Internet of Things (IoT) devices \cite{IoT:2014}; statistical data about cities and its population; and data from location-based social networks (LBSN)~\cite{Zheng2011,Zheng2012,traynor2012location}. As fog computing provides a scattered set of fog nodes throughout the urban environment, real-time identification of 
issues can be supported. For example, urban surveillance can be supported by fog computing to automate and improve public safety, supporting requirements of real-time
information processing and decision making~\cite{chen2016dynamic}.

\begin{figure}[!htbp]
\centering
        \includegraphics[width=.80\textwidth]{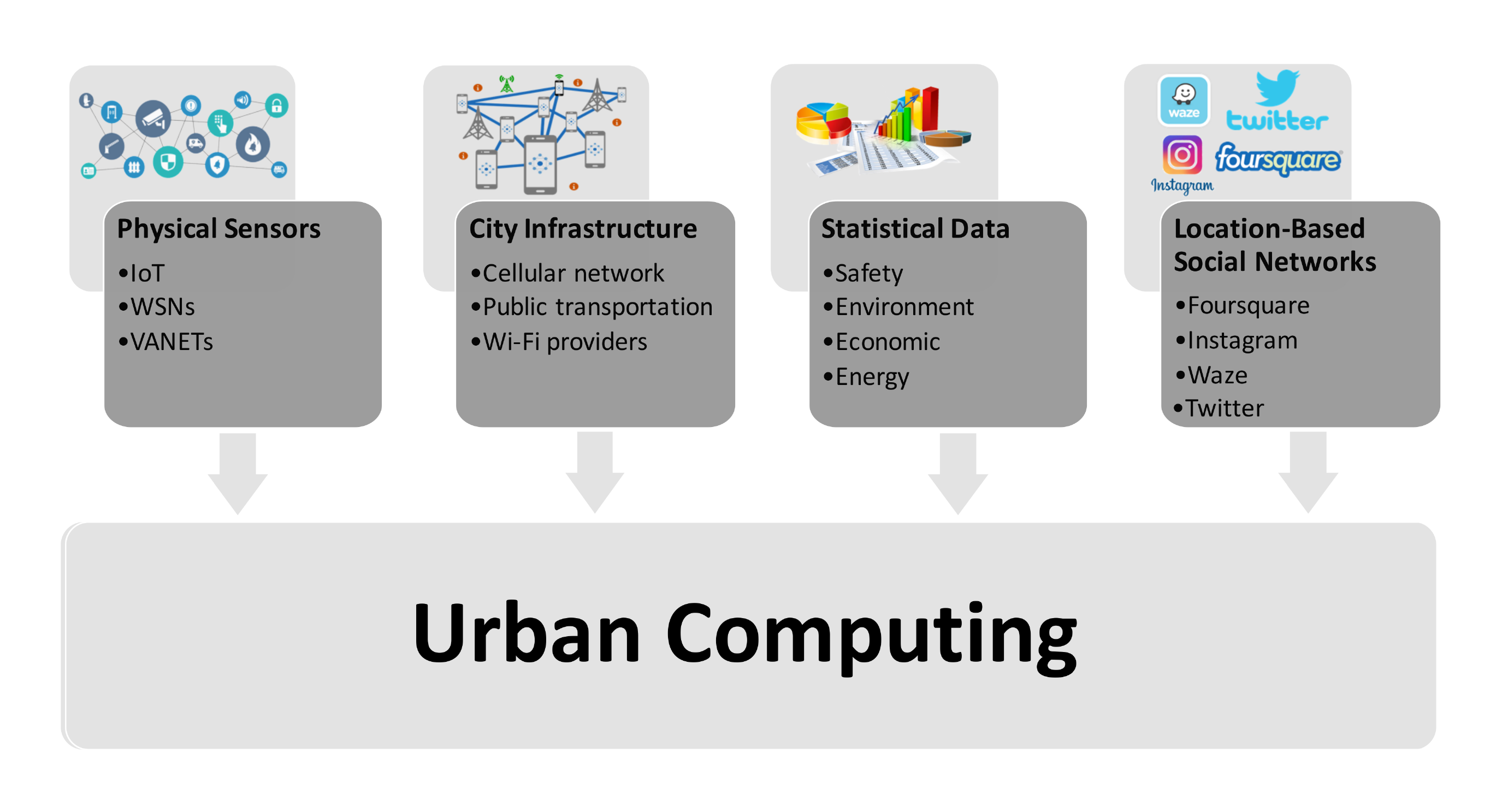}
\caption{Typical urban data sources.}\label{figUrbanDataSources}
\end{figure}

Urban computing aims to understand the aspects of the urban phenomena and also provide estimates about the future of cities. It is an interdisciplinary area, and in the context of computer science, urban computing intersects with sensor networks, computer networks, vehicles networks, social networks, distributed systems, and artificial intelligence. As urban computing is fairly comprehensive, a possible way of classifying research efforts in this area is through the data considered. Figure \ref{figUrbanDataSources} illustrates the main data sources used by studies in the area of urban computing. Most typical urban data sources can utilize fog nodes to process and transfer data between them or to the cloud for long-term storage or further processing.

Each of these sources, shown in Figure \ref{figUrbanDataSources}, is briefly described below:

\begin{itemize}

\item \textbf{Physical sensors:} They provide data that is obtained through the installation of sensors dedicated to certain applications, for example, inductive-loop traffic detectors to detect the volume of traffic in streets, sensors for monitoring air quality in various parts of the city, sensors for monitoring noise levels, and sensors in vehicles. One problem with the physical sensor data source is the difficulty in obtaining the data. In addition, there is a considerable cost for building a sensor network, when it is needed, and, generally, the deployment of sensors in the city demands special authorizations from the city hall. Besides, when it is desired to build a vehicular network, permissions and adaptations of vehicles of users are necessary, which could be troublesome. 

\item \textbf{Statistical data:} It consists of data related to a statistical study on a specific population, e.g., its demography, its health, and its social aspects. In addition, data on urban dynamics, such as economic, e.g., stock prices and housing prices; environment, e.g., flooding occurrences or agriculture details; safety, e.g., crimes committed and prisons made; and energy, e.g., gas consumption and electricity demand. It is possible to find multiple data sources on the Web from this category to some cities, and, typically, these data are open and easy to obtain. This type of data source is gaining popularity, particularly after government initiatives related to open data. However, these data may not be always available for the location we may intend to study. Another difficulty is the diversity of formats in which the data are available, for instance in tables, maps, graphs, calendars, forms, among others~\cite{barbosa2014structured}.

\item \textbf{Infrastructure of cities:} It provides data that is captured by taking advantage of existing city infrastructures that are created for other purposes. This includes cellular telephone networks. Cell phone signals from a large group of people have been used to characterize and predict individual's mobility and, consequently, to improve urban planning \cite{MucelliRezendeOliveira2017176,naboulsi2014classifying}. Other examples of city infrastructures able to provide usage data include WiFi service providers or public transportation systems. In particular, in this latter, it is very common the use of RFID cards to record users' bus and subway usage. Nevertheless, the difficulty here is that, typically, only the city or specific companies have access to this type of data.

\item \textbf{Location-based social networks (LBSNs):} They are systems that combine online social networks features and also allow users to share data containing spatio-temporal information. Location-based social networks provide urban data that implicitly have social aspects, such as user's preferences and routines \cite{Zheng2011,Zheng2012,traynor2012location}. This is due to the active and voluntary user participation, acting as a sort of social sensor, in a distributed process of sharing personal and also data about various aspects of the city in Web services. One key point is that users in these systems can manually determine when, how, where, and what to share.

LBSNs became quite popular partially due to the increased use of mobile devices, such as smartphones and tablets. These devices typically contain several sensors, e.g., GPS and accelerometer, enabling users to explore them to sense the environment, and, with that, having the opportunity to enrich LBSN data. LBSNs provide a new avenue of opportunities to access data on a global scale. 

There are several examples of location-based social networks already deployed on the Internet, such as (1) Foursquare, with more than $50$ million users monthly using it \cite{FourquareUsers}, which allows users to share locations they are visiting with their friends; (2) Waze\footnote{https://waze.com.}, with $65$ million active monthly users \cite{wazeUsers}, which serves to report traffic conditions in real-time; and (3) Instagram\footnote{https://instagram.com.}, a company with $700$ million monthly active users in 2017 \cite{instagramUsers}, which allows users to send real-time images to the system. Another example of LBSN is Twitter\footnote{https://twitter.com.}, a system with about $313$ million monthly active users in 2016 \cite{twitterUsers}, which allows its users to share personal updates as short text messages with up to 140 characters, known as ``tweets''.  Data from all those systems allow us to monitor various aspects of cities in near real-time, to which fog computing can be handy for analytics~\cite{Bonomi2014}. Taking as an example this former system, people could use their portable devices to share tweets containing real-time information about demonstrations or accidents in the city, allowing, for instance, unexpected problems to be identified by city authorities in real-time using fog computing nodes within the city boundaries, as for example for detection of various types of events that could threaten the pipelines integrity in the city~\cite{tang2015hierarchical}.

\end{itemize}

%Mobility and Smart Cities (Edmundo Madeira)
% (about 1.5 page per topic, including figures,emphasizing main contributions in the literature)

\subsubsection{Mobile Applications}
\label{sec:mobility}
        As electronic devices in general get smaller and smarter, they get embedded in
virtually all kinds of IoT objects. With more and more such devices being carried
by people and onboard vehicles (e.g., cars, bicycles, motorcycles, trains, and so on),
data generation and consumption also acquire mobile characteristics: data sources
and consumers can move around and change the network topology as well as data
traffic patterns. As a consequence, the need for computing power dynamically
changes in the geographical dimension, following the mobility of devices at the
edge of the network.
    
    The increase in the variety of connected devices also brought an increase in
applications that run on those devices. On the other hand, mobile devices, whether in
vehicles or not (e.g., wearable~\cite{mukhopadhyay2015wearable} or in-body
devices~\cite{kiourti2017review}), often have reduced computing capacity or
power constraints. Therefore, offloading has been studied to help in reducing
energy consumption \cite{kumar2010cloud,you2017energy} and response
time~\cite{kosta2012thinkair,kumar2013survey} of applications.
    
    Offloading from mobile devices to the cloud can help in saving a mobile device's
battery and also in reducing processing time for applications. Fog computing
adds computing layers between users and cloud, which can be also used to
offload~\cite{yi2015fog,shi2016edge}; applications can be developed to take
take offloading into consideration~\cite{orsini2015computing}. Mobility at the
edge and scattered fog resources throughout a hierarchy of computing power
introduces new variables for proper resource management: ideally, to maintain
latency and quality of service at the best possible level, offloaded computing
and data should follow their users in the fog environment~\cite{7399400}.
    
    In cloud computing, virtual machine migration is used to balance data center
load and also to consolidate virtual machines in fewer servers. In fog
computing, virtual machine migration can be utilized to replicate or move users
data and computing along their paths within a city, for example~\cite{7424534}.
However, proper resource management for multi-tiered fog environments is still
a challenge~\cite{mahmud2018fog}.
    
    Different mobility scenarios with heterogeneous applications can be supported by
the fog. In current societies, human mobility can be consistently predicted most
of the time~\cite{song2010limits}. Therefore, offloading and replicating data
and respective computing can take advantage of a predicted route to prevent
delays to move data during application needs. On the other hand, errors in
predictions or unpredictable patterns due to lack of history data also occur,
and must be properly addressed~\cite{diogo2018}.

Some mobile applications that perform offloading, such as assisted driving and health monitoring, require resiliency in the communication between mobile devices and fog/cloud. Path splitting and multi-path routing strategies can be applied to provide a certain level of resilience \cite{Velasquez2018}. However, device mobility brings new challenges, in the sense that a mobile device may be connected to different fogs along its path.

In some types of networks with mobile users, such as vehicular and cellular networks, offloading computation can be performed to another node of the same network, directly or via a relay. For such, D2D communication is used. For example, in a vehicular network, nearby vehicles may have idle computing resources. In this case, a task of a vehicle can be subdivided into smaller subtasks and the subtasks can opportunistically be offloaded to the neighbor vehicles via V2V communication~\cite{8300317}. Moreover, nearby vehicles can be used to deliver data and computation to/from the fog/cloud hierarchy when necessary.

%Industry 4.0 (Rizos Sakellariou)
% (about 1.5 page per topic, including figures,emphasizing main contributions in the literature)

\subsubsection{The Industrial Internet of Things}

The use of IoT technologies in industrial settings has been hailed as a highly innovative application with great potential to transform industry and manufacturing that may lead to a new industrial revolution, often coined with the term Industry 4.0 \cite{6714496,doi:10.1080/00207543.2017.1308576}. The vision is that the plethora of data that can be collected at all stages of production may form the basis for increased digitization leading to innovative processes, services and products of increased business value \cite{Kagermann2015}. This has resulted in lots of momentum and investment in projects, research and standardization efforts built around the so-called Industrial Internet of Things \cite{Serpanos2018,Jeschke2017}.

Naturally, the Industrial Internet of Things can be linked with Cloud and Fog Computing infrastructures that manage in efficient and effective ways the large amount of data that can potentially be generated. Lots of work has already discussed different aspects of this interaction
\cite{7328144,HOSSAIN2016192,Steiner2016}. Among topics that have attracted significant attention it is worth mentioning the standardization effort towards a reference architecture proposed by the the Industrial Internet Consortium \cite{b12}. A variety of system architecture aspects for Industry 4.0 ecosystems that are built upon the Industrial Internet of Things are discussed in the literature; some indicative work can be found in \cite{7785890,7467436,smartcomp,es2018}. These aspects will need to be enhanced as IoT-Fog-Cloud ecosystems become common place.
In relation to specific issues of increasing research interest one can highlight  the body of work on security 
\cite{10.1007/978-981-13-0311-1_30,7445139}, networking and communication \cite{7883994}, as well as data management
\cite{MOURTZIS2016290,TAO2018,8259028,es2018b,alma9956512960001631}.

%%%%%%%%%%%%%
% Future directions %%%
%%%%%%%%%%%%%
\section{Future directions}
\label{sec:challenges}
In this section we present several future directions for further research development in scenarios combining IoT, fog and cloud computing.
%Same topics from Literature review -- a couple of paragraphs to half page per topic.

%Same topics from Literature review -- a couple of paragraphs to half page per topic.

\subsection{Fog and 5G for IoT}
While the first 5G deployments are expected in the next couple of years, several challenges remain in how these deployments will support IoT services integrated with cloud and fog computing. Some of those challenges are outlined below.

In 5G, to realize the idea of network slicing in support of a set of services with specific performance requirements will require end-to-end resource management across wireless, optical, packet, fog nodes, and cloud domains. Recent advances in network virtualization provide a roadmap for this, but they have not yet achieved integrated orchestration of resources across all those domains. While slicing is, as mentioned previously, a key expected feature of 5G, it is unlikely that it will be fully realized in the initial deployments of the technology. 

Another requirement is the development of middleware and APIs that become de facto standards to communicate device requirements and capabilities to the network, and network conditions and feasible quality of service guarantees to the devices. This is needed for the fine-grained resource allocation to different network services, avoiding over- and under-provisioning including the fog resources, and for the automated establishment of service level agreements between an IoT service and the network or a slice.

In this context, it is necessary to devise efficient management mechanisms for increasingly heterogeneous and complex networks that adopt diverse wireless technologies (for IoT, those include LoRAWAN, Sixfox, and NB-IoT) and that comprise multiple models of ownership of networked resources from the edge (devices and fog) up to the cloud. Technology adoption and success, as always, will also depend on the development and maturity of business models for IoT services, remembering that the challenges are not only technical and also involve matters of public policy and investment decisions by operators and service providers.
%Same topics from Literature review -- a couple of paragraphs to half page per topic.

\subsection{Serverless Computing}
    Microservices management throughout the IoT-Fog-Cloud hierarchy presents challenges associated to the movement of services among IoT, fog, and cloud devices. The automatic adaptation of the execution of microservices must consider deployment location and context, but should also not neglect resource constraints that may exist at each level of the fog. To achieve this automatic and transparent adaptation, services reconfiguration that consider quality of service requirements is a challenge, where a service ranking approach can be implemented, for instance, to help multi-criteria decision making during reconfiguration.
    
    The heterogeneity of network across the IoT-Fog-Cloud ecosystem is also challenging for microservices deployment and reconfiguration. Standalone services can have network requirements to the data sources, which can be achieved through network technologies such as network virtualization and software defined networks (SDN). In this case, the need for reconfiguration of services includes a reconfiguration of the network to ensure requirements will remain in place. On the other hand, composition of services with different requirements can also be enacted vertically in the hierarchy, where a reconfiguration of services (and network, if necessary) is even more complex due to services heterogeneity in terms of computing needs and requirements (e.g., latency).

\subsection{Resource Allocation and Optimization}

	Optimization in resource allocation becomes more challenging as the number of variables increase as well as when these variables change more often over time. The composition of devices in the IoT-Fog-Cloud continuum brings new variables as the heterogeneity of devices and applications reach unprecedented levels. Moreover, network topology is expected to constantly change with device mobility and variable application requirements, introducing a more dynamic behavior to the system. This dynamic nature of the system along with high levels of heterogeneity call for dynamic, multi-criteria resource allocation strategies that can cope with the constantly changing environment. Resource management systems and multi-criteria schedulers that can rapidly optimize resource allocation in face of such changes are challenging, as the number of variables can exponentially expand the search space leading to long scheduler execution times. A trade-off between scheduler optimality and decision making turnaround time should ideally depend on user and application requirements, such as deadlines and acceptable delays. A parametrized scheduler to satisfactorily weigh such trade-offs in the IoT-Fog-Cloud continuum is yet to be modeled and developed.
    
    In parallel with the dynamic and heterogeneous scenario above, IoT applications often rely on data streams, which means the volume and velocity of data is an important input to the resource allocation decision. While in job-based systems the job's input data is usually measured in size, when stream processing (or complex event processing) takes place, processing requirements are based on the operation over the data stream and the frequency data is collected and streamed. As a consequence, schedulers are not aware of the whole optimization problem beforehand, and, thus, online optimization schemes would be more suitable to adapt the resource allocation over time.

%Same topics from Literature review -- a couple of paragraphs to half page per topic.

\subsection{Energy Consumption}

The proliferation of IoT devices and the ever increasing rate of data produced are increasing pressures on energy consumption. One should expect that such pressures will have to be addressed at both hardware and software levels as well as their interplay. Among the various approaches for energy efficient hardware design, approximate computing seems an interesting approach, not only at the hardware level \cite{7428027}. In terms of software, extensive work will need be carried out to take into account energy profiling characteristics of devices, infrastructures and applications. Different trade-offs will need to be studied and exploited: sacrificing some level of performance for significant energy savings may be an acceptable trade-off in many circumstances. 

An important direction for future research in minimizing energy consumption should focus on examining in more detail the role and impact of data in the IoT-Cloud-Fog ecosystem, along the lines of what has been termed as `economical data management' \cite{samos2018}. The idea should be to examine in detail the importance of different types of data and whether all data is needed all the time. This requires detailed assessments of how often it may be necessary to generate, transfer, store or process all different types of data. By associating different data management strategies with their corresponding energy consumption cost, the objective should be to find Pareto-optimal solutions. In this way, besides avoiding non-optimal solutions, applications can operate adaptively and choose appropriate trade-offs lying on the Pareto front according to user or system requirements. This type of research will need significant work in building and linking appropriate energy consumption models for all different components of an IoT-Fog-Cloud ecosystem. 
%Same topics from Literature review -- a couple of paragraphs to half page per topic.

%\paragraph{\textbf{Data management and locality}}:\\[5pt]
\subsection{Data Management and Locality}
There are several open issues related to data management and locality in IoT-Fog-Cloud computing systems.
First and foremost, these systems are typically composed of a broad set of heterogeneous communication technologies such as cellular, wireless, wired, and radio frequency.
This means that the system’s orchestration has to be able to handle distinct underlying networks as well as different addressing schemes.
Centralizing all the resources within the cloud partially solves some issues, like availability, scalability, and interoperability, however, it introduces new ones, e.g., network congestion and higher latency, which can be mitigated with fog and edge computing.
One issue is how to measure and quantify the trade-off between placing data and services at the cloud or fog level. 

A common approach to improve on this issue is through smart service placement. 
In this way, it is possible to provide data locality by placing the services needed close to the data that it operates on. 
However, one of the open issues here is how to chose the services that are going to be placed at the edge nodes and for how long.
Applications that do not require high-processing power and need to analyze large chunks of data are good candidates. 
On the other hand, several interactive applications, such as augmented reality, may require high-processing power and ultra-low latency times, so they are also good candidates.
Because of that, choosing the best candidates is a complex task.
To make matters worse, if the diversity of the data that have to be transmitted or analyzed and the multitude of communication technologies are considered, the problem becomes more complex calling for sophisticated multi-criteria optimization strategies to be developed. 

%Same topics from Literature review -- a couple of paragraphs to half page per topic.
\subsection{Applying Federation Concepts to Fog Computing and IoT}
Federations will be widely used in many different application domains.
The outstanding challenge here is how can federation capabilities be
best applied in fog and IoT environments?
The easiest answer is to simplify the deployment and governance models
to be used.
This can be done by relying on out-of-band information as much as
possible.
Fog/IoT federation can be simplified if a particular federation
involves only a small number of known, fixed IoT device types.
This may also only require a small set of known, fixed roles or
attributes to manage the acquired data.
It may also be possible to use simple hardware-based methods to
establish fog node identity.

The more general question is could a Federation Manager be devised that
is tailored for fog and IoT environments?
Standardizing such a federation profile would enable the wider
deployment and use of federations in such domains.
Scalability will always be a concern as the number of fog and IoT
devices being managed within one federated environment increases.
Any kind of Fog/IoT Federation Manager would have to be designed to
cope with scalability requirements.

\subsection{Trust Models to Support Federation in Fog and IoT Environments}

Identity and trust are the cornerstones of federation management.
While a number of methods exist for establishing identity and trust,
the only feasible methods are based on cryptographic methods.
An inherent property of IoT environments, though, is that the closer
to the {\em edge} one gets, the more resource-constrained the devices
will become.
This means the use of cryptographic methods to manage federations will
have to stop short of the IoT devices themselves.
Being able to support cryptographic operations will thus be a
distinguishing feature of fog nodes.
This poses the question how lightweight can cryptographic methods be
made such that federations can be supported on less powerful fog
nodes, and deployed closer to the IoT devices themselves.
This is an outstanding challenge for establishing identity and trust
to support federations in Fog/IoT environments.

%Same topics from Literature review -- a couple of paragraphs to half page per topic.

\subsection{Orchestration in Fog for IoT}
Despite recent developments in the area of fog orchestration for the Internet of Things, there are still several open issues that need to be addressed. 

First and foremost, privacy must be tackled in accordance to the European Union General Data Protection Regulation as well as similar regulations being enforced worldwide. This is an important issue, since the fog nodes, being placed close to the end users gather, store and process data that may potentially be used to violate users privacy. The different security perspectives of the fog-based IoT environment are also extremely important given its distributed, dynamic and large-scale nature. In particular, security mechanisms must be developed that prevent software, hardware or network attacks to fog orchestrator nodes. 

Performance of fog orchestration for the IoT faces several challenges, in particular within the context of 5G networks. The high density of devices together with the latency and reliability requirements of critical applications as well as node mobility, raise important issues concerning the monitoring of the whole system, which is fundamental for a proper resource management. Component selection and placement are also essential aspects that directly affect performance of dynamic fog orchestration and need to be explored in the future along with research on efficient mechanisms to prevent overloading and avoid orchestration delays.

Considering the large amounts of multi-dimensional data in fog-based IoT scenarios, approaches that provide multiple levels of real-time data analytics together with efficient optimization mechanisms ought to be researched. One important characteristic that must drive this line of research is the layered structure under the control of the fog orchestrators, which requires the development of cross-layer solutions.

All these perspectives have been identified by the OpenFog consortium and drive ongoing and future research in the area of fog orchestration for the IoT.

%Same topics from Literature review -- a couple of paragraphs to half page per topic.

\subsection{Business and Service Models}
	While cloud computing has been offering a variety of business and service models through the years, it is not clear yet if fog computing can simply incorporate the cloud models or if new business or service models would be feasible. 
    
    The cloud established way of charging and billing is suitable for a variety of computing services. On the other hand, fog infrastructure management can involve a wider set of stakeholders, including autonomous systems within academia, industry, offices, small- and mid-sized businesses, telecom operators, public authorities, and so on. Currently, the fog can be deployed as a hybrid cloud, where local resources (e.g., a small private cloud) are extended with resources from the cloud. When other players are introduced in the hierarchy from IoT to the cloud, this involves a set of devices that are managed by different entities (e.g., IoT devices can be owned by the state while fog nodes by a cloud company; or the opposite). How services for IoT combined with services from fog and cloud computing can be offered, monitored and charged can be challenging when multiple players at different levels and with heterogeneous interests are involved.

%Same topics from Literature review -- a couple of paragraphs to half page per topic.

\subsection{Mobility}
        Efficiently allocating resources for mobile users is a challenge in fog computing.
Users and devices mobility patterns are an important aspect to provide proper
service when offloading to the fog occurs. Dealing with a large set of mobile users
with diverse applications and requirements is a highly dynamic
scenario, which makes resource management challenging. Sets of cloudlets can be
overloaded during certain periods of time, when many users are around a specific
location (e.g., a city center in busy times of the day), requiring resource
management entities to allocate more distant resources for some applications and
users. Such decision making needs information about users mobility patterns and
their application requirements and/or usage patterns to result in an allocation of
fog resources that maximize the user's satisfaction (applications quality of
service or users quality of experience).
    
    The hierarchy of computing brought by the fog makes the resource management
challenge different from cloud computing, content delivery networks, or other
mobile computing infrastructures on the edge. Besides deciding where to place
data and computing of each mobile user, the speed of each user may also play a
role in the decision: for example, higher speed users could have their data
placed in cloudlets at a higher level in the fog hierarchy to minimize the
amount of migrations needed, also reducing network utilization and
unavailability during migration times. When relay and multi-hop communication among mobile devices is added to communicate with the fog/cloud hierarchy, the decision-making on resource allocation is even more challenging.
    
    The aggregation of user mobility, fog/cloud hierarchy, and application requirements into a resource allocation model is a challenge yet to be addressed.

	%Same topics from Literature review -- a couple of paragraphs to half page per topic.

\subsection{Urban Computing}
Although several research efforts related to urban computing have been performed recently, it is possible to find open issues and opportunities for studying cities and societies using LBSN data. Several previous studies model LBSN data as static structures, not taking into account the temporal dynamics. Even though this is an accepted strategy, this representation might result in loss of relevant information in certain cases. In addition, another example of the challenge is to work with a large number of data that LBSNs can potentially provide. This imposes several challenges related to, for example, processing, storage, and indexing in real-time when using tools of conventional data processing systems and database management. One possible direction is to extend cloud-based Complex Event Processing~\cite{8029323} to be also deployed in fog nodes. In addition, LBSN data exploration may threaten the privacy of users. For example, LBSN data could be explored to deduce users' preferences and particular behavior. With this, users have no guarantee that their private life will not be violated by others. It is a challenge to ensure people's privacy while relying on data that can be potentially sensitive, but a geographically constrained fog computing within city boundaries might be developed to handle sensitive data from citizens.

%Same topics from Literature review -- a couple of paragraphs to half page per topic.

\subsection{The Industrial Internet of Things}

Designing software that exploits the Industrial Internet of Things constitutes a ``system of systems'' challenge. Taking into account the whole Iot-Fog-Cloud continuum, addressing the complexity of this challenge will require frameworks that enable interoperability but are also able to cope with varying and possibly conflicting user and system requirements. It can be envisaged that not a single framework would be able to cope with all possible scenarios. What becomes apparent is that the traditional, rather centralized approach to organize and handle data in industrial settings would have to change. Decentralized approaches may become more common place and different levels of importance, on different occasions, may be associated to subsets of the IoT-Fog-Cloud data. Handling such dynamically changing requirements on data, services and processes, at the same time respecting various operational constraints and goals, can be a major challenge. Finally, security aspects, often mentioned as key issues to safeguard the integrity of the Industrial Internet of Things \cite{7167238}, will need to be considered extensively.

%%%%%%%%%%%%%
% Conclusion %%%
%%%%%%%%%%%%%
% \begin{table}[]
% \begin{tabular}{llll}
% Topic               & Challenges & Paper & Problem addressed \\
% Resource allocation &            &       &                   \\
% Orchestration       &            &       &                   \\
% Services            &            &       &                   \\
% Energy consumption  &            &       &                   \\
% Federation          &            &       &                   \\
% Data locality       &            &       &                   \\
% Urban computing     &            &       &                   \\
% Mobile              &            &       &                   \\
% Industrial IoT      &            &       &                  
% \end{tabular}
% \end{table}

\section{Conclusion}
	The expansion of the Internet of Things demands new paradigms for data collection and processing. Fog computing has emerged as one way of dealing with the big data resulting from IoT. The combination of fog and cloud computing is a promising way of providing full capabilities to support IoT and its wide range of requirements, from low-latency/real-time to processing- or storage-demanding applications.
    
	New applications developed as a result from the IoT expansion call for location awareness, low latency, and mobility support in a geo-distributed scenario. This paper defined and discussed key aspects and distinct scenarios of edge and fog computing as well as how they can extend and complement the already established cloud environment to support IoT applications. Several aspects of fog and cloud computing have already been addressed in the literature; this paper has discussed how some of them are still challenging to provide an effective infrastructure for IoT data processing and storage.
    
    As fog computing research evolves with IoT and some of their challenges are addressed, we expect new challenges to arise in terms of resource management and its efficiency as the amount of devices and heterogeneous applications keep growing.

\label{sec:con}

%%%%%%%%%%%%%
% Acknowledgements%%
%%%%%%%%%%%%%
 \section*{Acknowledgements}
 The authors would like to thank the following agencies for partially supporting this research: the European Commission H2020 programme under grant agreement no. 688941 (FUTEBOL), as well from the Brazilian Ministry of Science, Technology, Innovation, and Communication (MCTIC) through RNP and CTIC; the S\~{a}o Paulo Research Foundation (FAPESP), grants \#2015/16332-8, \#2018/02204-6, and \#2015/24494-8; the MobiWise project: from mobile sensing to mobility advising (P2020 SAICTPAC/0011/2015), co-financed by COMPETE 2020, Portugal 2020-POCI, European Regional Development Fund of European Union, and the Portuguese Foundation of Science and Technology; CAPES and CNPq.

%\section*{References}
%\bibliography{FogIoT}

\begin{thebibliography}{100}
\expandafter\ifx\csname url\endcsname\relax
  \def\url#1{\texttt{#1}}\fi
\expandafter\ifx\csname urlprefix\endcsname\relax\def\urlprefix{URL }\fi
\expandafter\ifx\csname href\endcsname\relax
  \def\href#1#2{#2} \def\path#1{#1}\fi

\bibitem{GUBBI20131645}
J.~Gubbi, R.~Buyya, S.~Marusic, M.~Palaniswami, {Internet of Things (IoT)}: A
  vision, architectural elements, and future directions, Future Generation
  Computer Systems 29~(7) (2013) 1645 -- 1660.
\newblock \href
  {http://dx.doi.org/https://doi.org/10.1016/j.future.2013.01.010}
  {\path{doi:https://doi.org/10.1016/j.future.2013.01.010}}.

\bibitem{7912261}
L.~F. Bittencourt, J.~Diaz-Montes, R.~Buyya, O.~F. Rana, M.~Parashar,
  Mobility-aware application scheduling in fog computing, IEEE Cloud Computing
  4~(2) (2017) 26--35.

\bibitem{dinh2013survey}
H.~T. Dinh, C.~Lee, D.~Niyato, P.~Wang, A survey of mobile cloud computing:
  architecture, applications, and approaches, Wireless communications and
  mobile computing 13~(18) (2013) 1587--1611.

\bibitem{shi2016edge}
W.~Shi, J.~Cao, Q.~Zhang, Y.~Li, L.~Xu, Edge computing: Vision and challenges,
  IEEE Internet of Things Journal 3~(5) (2016) 637--646.

\bibitem{Bonomi:2012:FCR:2342509.2342513}
F.~Bonomi, R.~Milito, J.~Zhu, S.~Addepalli, Fog computing and its role in the
  internet of things, in: Proceedings of the First Edition of the MCC Workshop
  on Mobile Cloud Computing, MCC '12, ACM, New York, NY, USA, 2012, pp. 13--16.
\newblock \href {http://dx.doi.org/10.1145/2342509.2342513}
  {\path{doi:10.1145/2342509.2342513}}.

\bibitem{sundmaeker2010vision}
H.~Sundmaeker, P.~Guillemin, P.~Friess, S.~Woelffl{\'e}, Vision and challenges
  for realising the internet of things, Cluster of European Research Projects
  on the Internet of Things, European Commision 3~(3) (2010) 34--36.

\bibitem{Armbrust:2010:VCC:1721654.1721672}
M.~Armbrust, A.~Fox, R.~Griffith, A.~D. Joseph, R.~Katz, A.~Konwinski, G.~Lee,
  D.~Patterson, A.~Rabkin, I.~Stoica, M.~Zaharia, A view of cloud computing,
  Commun. ACM 53~(4) (2010) 50--58.
\newblock \href {http://dx.doi.org/10.1145/1721654.1721672}
  {\path{doi:10.1145/1721654.1721672}}.

\bibitem{7214098}
Y.~Duan, G.~Fu, N.~Zhou, X.~Sun, N.~C. Narendra, B.~Hu, Everything as a service
  (xaas) on the cloud: Origins, current and future trends, in: 2015 IEEE 8th
  International Conference on Cloud Computing, 2015, pp. 621--628.

\bibitem{6295710}
L.~F. Bittencourt, E.~R.~M. Madeira, N.~L. S.~D. Fonseca, Scheduling in hybrid
  clouds, IEEE Communications Magazine 50~(9) (2012) 42--47.

\bibitem{mell2011nist}
P.~Mell, T.~Grance, et~al., The nist definition of cloud computing.

\bibitem{foster2001anatomy}
I.~Foster, C.~Kesselman, S.~Tuecke, The anatomy of the grid: Enabling scalable
  virtual organizations, The International Journal of High Performance
  Computing Applications 15~(3) (2001) 200--222.

\bibitem{ETSIMagazine2017}
{European Telecommunications Standards Institute (ETSI)}, {The Standard, News
  From ETSI}, ETSI Magazine 2.

\bibitem{openfog2017openfog}
O.~C. A.~W. Group, et~al., Openfog reference architecture for fog computing,
  OPFRA001 20817 (2017) 162.

\bibitem{8240187}
J.~C. Guevara, L.~F. Bittencourt, N.~L.~S. da~Fonseca, Class of service in fog
  computing, in: 2017 IEEE 9th Latin-American Conference on Communications
  (LATINCOM), 2017, pp. 1--6.

\bibitem{Al-Fares:2008:SCD:1402946.1402967}
M.~Al-Fares, A.~Loukissas, A.~Vahdat, A scalable, commodity data center network
  architecture, SIGCOMM Comput. Commun. Rev. 38~(4) (2008) 63--74.
\newblock \href {http://dx.doi.org/10.1145/1402946.1402967}
  {\path{doi:10.1145/1402946.1402967}}.

\bibitem{Greenberg:2009:VSF:1594977.1592576}
A.~Greenberg, J.~R. Hamilton, N.~Jain, S.~Kandula, C.~Kim, P.~Lahiri, D.~A.
  Maltz, P.~Patel, S.~Sengupta, Vl2: A scalable and flexible data center
  network, SIGCOMM Comput. Commun. Rev. 39~(4) (2009) 51--62.
\newblock \href {http://dx.doi.org/10.1145/1594977.1592576}
  {\path{doi:10.1145/1594977.1592576}}.

\bibitem{7424534}
L.~F. Bittencourt, M.~M. Lopes, I.~Petri, O.~F. Rana, Towards virtual machine
  migration in fog computing, in: 2015 10th International Conference on P2P,
  Parallel, Grid, Cloud and Internet Computing (3PGCIC), 2015, pp. 1--8.

\bibitem{yi2015fog}
S.~Yi, Z.~Hao, Z.~Qin, Q.~Li, Fog computing: Platform and applications, in:
  2015 Third IEEE Workshop on Hot Topics in Web Systems and Technologies
  (HotWeb), IEEE, 2015, pp. 73--78.

\bibitem{7020884}
I.~Stojmenovic, Fog computing: A cloud to the ground support for smart things
  and machine-to-machine networks, in: 2014 Australasian Telecommunication
  Networks and Applications Conference (ATNAC), 2014, pp. 117--122.

\bibitem{Singh:2017:RRT:3147213.3147216}
A.~Singh, N.~Auluck, O.~Rana, A.~Jones, S.~Nepal, Rt-sane: Real time security
  aware scheduling on the network edge, in: Proceedings of the10th
  International Conference on Utility and Cloud Computing, UCC '17, ACM, New
  York, NY, USA, 2017, pp. 131--140.
\newblock \href {http://dx.doi.org/10.1145/3147213.3147216}
  {\path{doi:10.1145/3147213.3147216}}.

\bibitem{7098039}
M.~Aazam, E.~N. Huh, Fog computing micro datacenter based dynamic resource
  estimation and pricing model for iot, in: 2015 IEEE 29th International
  Conference on Advanced Information Networking and Applications, 2015, pp.
  687--694.

\bibitem{8004151}
R.~Vilalta, V.~Lopez, A.~Giorgetti, S.~Peng, V.~Orsini, L.~Velasco,
  R.~Serral-Gracia, D.~Morris, S.~D. Fina, F.~Cugini, P.~Castoldi, A.~Mayoral,
  R.~Casellas, R.~Martinez, C.~Verikoukis, R.~Munoz, Telcofog: A unified
  flexible fog and cloud computing architecture for 5g networks, IEEE
  Communications Magazine 55~(8) (2017) 36--43.

\bibitem{7987464}
M.~Taneja, A.~Davy, Resource aware placement of iot application modules in
  fog-cloud computing paradigm, in: 2017 IFIP/IEEE Symposium on Integrated
  Network and Service Management (IM), 2017, pp. 1222--1228.

\bibitem{7511465}
V.~B.~C. Souza, W.~Ramírez, X.~Masip-Bruin, E.~Marín-Tordera, G.~Ren,
  G.~Tashakor, Handling service allocation in combined fog-cloud scenarios, in:
  2016 IEEE International Conference on Communications (ICC), 2016, pp. 1--5.

\bibitem{6984239}
M.~Aazam, E.~N. Huh, Fog computing and smart gateway based communication for
  cloud of things, in: 2014 International Conference on Future Internet of
  Things and Cloud, 2014, pp. 464--470.

\bibitem{Hong2013}
K.~Hong, D.~Lillethun, U.~Ramachandran, B.~Ottenw\"{a}lder, B.~Koldehofe,
  Mobile fog: A programming model for large-scale applications on the internet
  of things, in: Proceedings of the Second ACM SIGCOMM Workshop on Mobile Cloud
  Computing, MCC '13, ACM, New York, NY, USA, 2013, pp. 15--20.
\newblock \href {http://dx.doi.org/10.1145/2491266.2491270}
  {\path{doi:10.1145/2491266.2491270}}.

\bibitem{openfog2017}
O.~Consortium, et~al., Openfog reference architecture for fog computing,
  Architecture Working Group.

\bibitem{Chiang2016}
M.~Chiang, T.~Zhang, Fog and iot: An overview of research opportunities, IEEE
  Internet of Things Journal 3~(6) (2016) 854--864.

\bibitem{Tozlu2012}
S.~Tozlu, M.~Senel, W.~Mao, A.~Keshavarzian, Wi-fi enabled sensors for internet
  of things: A practical approach, IEEE Communications Magazine 50~(6) (2012)
  134--143.

\bibitem{Chang2014}
K.~Chang, Bluetooth: a viable solution for iot? [industry perspectives], IEEE
  Wireless Communications 21~(6) (2014) 6--7.

\bibitem{Gomez2010}
C.~Gomez, J.~Paradells, Wireless home automation networks: A survey of
  architectures and technologies, IEEE Communications Magazine 48~(6) (2010)
  92--101.

\bibitem{Lu2004}
G.~Lu, B.~Krishnamachari, C.~S. Raghavendra, Performance evaluation of the ieee
  802.15.4 mac for low-rate low-power wireless networks, in: IEEE International
  Conference on Performance, Computing, and Communications, 2004, 2004, pp.
  701--706.

\bibitem{Mulligan2007}
G.~Mulligan, The 6lowpan architecture, in: Proceedings of the 4th Workshop on
  Embedded Networked Sensors, EmNets '07, ACM, New York, NY, USA, 2007, pp.
  78--82.
\newblock \href {http://dx.doi.org/10.1145/1278972.1278992}
  {\path{doi:10.1145/1278972.1278992}}.

\bibitem{Bouaziz2016}
M.~Bouaziz, A.~Rachedi, A survey on mobility management protocols in wireless
  sensor networks based on 6lowpan technology, Computer Communications 74
  (2016) 3 -- 15, current and Future Architectures, Protocols, and Services for
  the Internet of Things.
\newblock \href
  {http://dx.doi.org/https://doi.org/10.1016/j.comcom.2014.10.004}
  {\path{doi:https://doi.org/10.1016/j.comcom.2014.10.004}}.

\bibitem{CUNHA2016}
F.~Cunha, L.~Villas, A.~Boukerche, G.~Maia, A.~Viana, R.~A. Mini, A.~A.
  Loureiro, Data communication in vanets: Protocols, applications and
  challenges, Ad Hoc Networks 44 (2016) 90 -- 103.
\newblock \href {http://dx.doi.org/https://doi.org/10.1016/j.adhoc.2016.02.017}
  {\path{doi:https://doi.org/10.1016/j.adhoc.2016.02.017}}.

\bibitem{Fan2014}
C.~Fan, S.~Huang, Y.~Lai, Privacy-enhanced data aggregation scheme against
  internal attackers in smart grid, IEEE Transactions on Industrial Informatics
  10~(1) (2014) 666--675.

\bibitem{Jin2016}
H.~Jin, L.~Su, H.~Xiao, K.~Nahrstedt, Inception: Incentivizing
  privacy-preserving data aggregation for mobile crowd sensing systems, in:
  Proceedings of the 17th ACM International Symposium on Mobile Ad Hoc
  Networking and Computing, MobiHoc '16, ACM, New York, NY, USA, 2016, pp.
  341--350.
\newblock \href {http://dx.doi.org/10.1145/2942358.2942375}
  {\path{doi:10.1145/2942358.2942375}}.

\bibitem{Villas2013}
L.~A. Villas, A.~Boukerche, H.~S. Ramos, H.~A. B.~F. de~Oliveira, R.~B.
  de~Araujo, A.~A.~F. Loureiro, Drina: A lightweight and reliable routing
  approach for in-network aggregation in wireless sensor networks, IEEE
  Transactions on Computers 62~(4) (2013) 676--689.

\bibitem{Xiang2013}
L.~Xiang, J.~Luo, C.~Rosenberg, Compressed data aggregation: Energy-efficient
  and high-fidelity data collection, IEEE/ACM Trans. Netw. 21~(6) (2013)
  1722--1735.
\newblock \href {http://dx.doi.org/10.1109/TNET.2012.2229716}
  {\path{doi:10.1109/TNET.2012.2229716}}.

\bibitem{Li2011}
H.~Li, K.~Lin, K.~Li, Energy-efficient and high-accuracy secure data
  aggregation in wireless sensor networks, Computer Communications 34~(4)
  (2011) 591 -- 597, special issue: Building Secure Parallel and Distributed
  Networks and Systems.
\newblock \href
  {http://dx.doi.org/https://doi.org/10.1016/j.comcom.2010.02.026}
  {\path{doi:https://doi.org/10.1016/j.comcom.2010.02.026}}.

\bibitem{Zhang2017}
H.~Zhang, N.~Liu, X.~Chu, K.~Long, A.~H. Aghvami, V.~C.~M. Leung, Network
  slicing based 5g and future mobile networks: Mobility, resource management,
  and challenges, IEEE Communications Magazine 55~(8) (2017) 138--145.

\bibitem{Samdanis2016}
K.~Samdanis, X.~Costa-Perez, V.~Sciancalepore, From network sharing to
  multi-tenancy: The 5g network slice broker, IEEE Communications Magazine
  54~(7) (2016) 32--39.

\bibitem{7495388}
S.~Kitanov, E.~Monteiro, T.~Janevski, 5g and the fog — survey of related
  technologies and research directions, in: 2016 18th Mediterranean
  Electrotechnical Conference (MELECON), 2016, pp. 1--6.

\bibitem{7901475}
Y.~Ku, D.~Lin, C.~Lee, P.~Hsieh, H.~Wei, C.~Chou, A.~Pang, 5g radio access
  network design with the fog paradigm: Confluence of communications and
  computing, IEEE Communications Magazine 55~(4) (2017) 46--52.

\bibitem{yannuzzi2014key}
M.~Yannuzzi, R.~Milito, R.~Serral-Graci{\`a}, D.~Montero, M.~Nemirovsky, Key
  ingredients in an iot recipe: Fog computing, cloud computing, and more fog
  computing, in: Computer Aided Modeling and Design of Communication Links and
  Networks (CAMAD), 2014 IEEE 19th International Workshop on, IEEE, 2014, pp.
  325--329.

\bibitem{7548876}
O.~Bibani, S.~Yangui, R.~H. Glitho, W.~Gaaloul, N.~B. Hadj-Alouane, M.~J.
  Morrow, P.~A. Polakos, A demo of a paas for iot applications provisioning in
  hybrid cloud/fog environment, in: 2016 IEEE International Symposium on Local
  and Metropolitan Area Networks (LANMAN), 2016, pp. 1--2.

\bibitem{8116434}
A.~P. Silva, B.~A. Abreu, E.~B. Silva, M.~Carvalho, M.~Nunes, M.~Marotta,
  A.~Hammad, C.~F.~M. Silva, J.~F.~N. Pinheiro, C.~B. Both, J.~M.
  Marquez-Barja, L.~A. DaSilva, Impact of fog and cloud computing on an iot
  service running over an optical/wireless network testbed, in: 2017 IEEE
  Conference on Computer Communications Workshops (INFOCOM WKSHPS), 2017, pp.
  535--540.

\bibitem{pinedo2016scheduling}
M.~L. Pinedo, Scheduling: theory, algorithms, and systems, Springer, 2016.

\bibitem{kan2012machine}
A.~R. Kan, Machine scheduling problems: classification, complexity and
  computations, Springer Science \& Business Media, 2012.

\bibitem{1558639}
J.~Blythe, S.~Jain, E.~Deelman, Y.~Gil, K.~Vahi, A.~Mandal, K.~Kennedy, Task
  scheduling strategies for workflow-based applications in grids, in: CCGrid
  2005. IEEE International Symposium on Cluster Computing and the Grid, 2005.,
  Vol.~2, 2005, pp. 759--767 Vol. 2.

\bibitem{meng2010improving}
X.~Meng, V.~Pappas, L.~Zhang, Improving the scalability of data center networks
  with traffic-aware virtual machine placement, in: INFOCOM, 2010 Proceedings
  IEEE, IEEE, 2010, pp. 1--9.

\bibitem{Pietri:2016:MVM:2988524.2983575}
I.~Pietri, R.~Sakellariou, Mapping virtual machines onto physical machines in
  cloud computing: A survey, ACM Computing Surveys 49~(3) (2016) 49:1--49:30.
\newblock \href {http://dx.doi.org/10.1145/2983575}
  {\path{doi:10.1145/2983575}}.

\bibitem{li2013energy}
X.~Li, Z.~Qian, S.~Lu, J.~Wu, Energy efficient virtual machine placement
  algorithm with balanced and improved resource utilization in a data center,
  Mathematical and Computer Modelling 58~(5-6) (2013) 1222--1235.

\bibitem{pandey2010particle}
S.~Pandey, L.~Wu, S.~M. Guru, R.~Buyya, A particle swarm optimization-based
  heuristic for scheduling workflow applications in cloud computing
  environments, in: Advanced information networking and applications (AINA),
  2010 24th IEEE international conference on, IEEE, 2010, pp. 400--407.

\bibitem{zeng2016joint}
D.~Zeng, L.~Gu, S.~Guo, Z.~Cheng, S.~Yu, Joint optimization of task scheduling
  and image placement in fog computing supported software-defined embedded
  system, IEEE Transactions on Computers 65~(12) (2016) 3702--3712.

\bibitem{7774691}
E.~d.~Lara, C.~S. Gomes, S.~Langridge, S.~H. Mortazavi, M.~Roodi, Poster
  abstract: Hierarchical serverless computing for the mobile edge, in: 2016
  IEEE/ACM Symposium on Edge Computing (SEC), 2016, pp. 109--110.

\bibitem{7802525}
M.~Villari, M.~Fazio, S.~Dustdar, O.~Rana, R.~Ranjan, Osmotic computing: A new
  paradigm for edge/cloud integration, IEEE Cloud Computing 3~(6) (2016)
  76--83.

\bibitem{7061425}
J.~Pan, R.~Jain, S.~Paul, T.~Vu, A.~Saifullah, M.~Sha, An internet of things
  framework for smart energy in buildings: Designs, prototype, and experiments,
  IEEE Internet of Things Journal 2~(6) (2015) 527--537.

\bibitem{7805265}
D.~Minoli, K.~Sohraby, B.~Occhiogrosso, Iot considerations, requirements, and
  architectures for smart buildings—energy optimization and next-generation
  building management systems, IEEE Internet of Things Journal 4~(1) (2017)
  269--283.

\bibitem{5940920}
V.~M. Rohokale, N.~R. Prasad, R.~Prasad, A cooperative internet of things (iot)
  for rural healthcare monitoring and control, in: 2011 2nd International
  Conference on Wireless Communication, Vehicular Technology, Information
  Theory and Aerospace Electronic Systems Technology (Wireless VITAE), 2011,
  pp. 1--6.

\bibitem{LEE2015431}
I.~Lee, K.~Lee, The internet of things (iot): Applications, investments, and
  challenges for enterprises, Business Horizons 58~(4) (2015) 431 -- 440.
\newblock \href
  {http://dx.doi.org/https://doi.org/10.1016/j.bushor.2015.03.008}
  {\path{doi:https://doi.org/10.1016/j.bushor.2015.03.008}}.

\bibitem{6291714}
V.~Hanumaiah, S.~Vrudhula, Energy-efficient operation of multicore processors
  by dvfs, task migration, and active cooling, IEEE Transactions on Computers
  63~(2) (2014) 349--360.

\bibitem{Dabbelt:2016:VPE:2934495.2934497}
D.~Dabbelt, C.~Schmidt, E.~Love, H.~Mao, S.~Karandikar, K.~Asanovic, Vector
  processors for energy-efficient embedded systems, in: Proceedings of the
  Third ACM International Workshop on Many-core Embedded Systems, MES '16, ACM,
  New York, NY, USA, 2016, pp. 10--16.
\newblock \href {http://dx.doi.org/10.1145/2934495.2934497}
  {\path{doi:10.1145/2934495.2934497}}.

\bibitem{7284406}
D.~Hackenberg, R.~Schöne, T.~Ilsche, D.~Molka, J.~Schuchart, R.~Geyer, An
  energy efficiency feature survey of the intel haswell processor, in: 2015
  IEEE International Parallel and Distributed Processing Symposium Workshop,
  2015, pp. 896--904.

\bibitem{7927716}
F.~Conti, R.~Schilling, P.~D. Schiavone, A.~Pullini, D.~Rossi, F.~K.
  Gürkaynak, M.~Muehlberghuber, M.~Gautschi, I.~Loi, G.~Haugou, S.~Mangard,
  L.~Benini, An iot endpoint system-on-chip for secure and energy-efficient
  near-sensor analytics, IEEE Transactions on Circuits and Systems I: Regular
  Papers 64~(9) (2017) 2481--2494.

\bibitem{Luo:2018:NDT:3195970.3196080}
S.~Luo, C.~Zhuo, H.~Gan, Noise-aware dvfs transition sequence optimization for
  battery-powered iot devices, in: Proceedings of the 55th Annual Design
  Automation Conference, DAC '18, ACM, New York, NY, USA, 2018, pp. 27:1--27:6.
\newblock \href {http://dx.doi.org/10.1145/3195970.3196080}
  {\path{doi:10.1145/3195970.3196080}}.

\bibitem{Urgaonkar:2011:OPC:1993744.1993766}
R.~Urgaonkar, B.~Urgaonkar, M.~J. Neely, A.~Sivasubramaniam, Optimal power cost
  management using stored energy in data centers, in: Proceedings of the ACM
  SIGMETRICS Joint International Conference on Measurement and Modeling of
  Computer Systems, SIGMETRICS '11, ACM, New York, NY, USA, 2011, pp. 221--232.
\newblock \href {http://dx.doi.org/10.1145/1993744.1993766}
  {\path{doi:10.1145/1993744.1993766}}.

\bibitem{BELOGLAZOV201147}
A.~Beloglazov, R.~Buyya, Y.~C. Lee, A.~Zomaya, Chapter 3 - a taxonomy and
  survey of energy-efficient data centers and cloud computing systems, Vol.~82
  of Advances in Computers, Elsevier, 2011, pp. 47 -- 111.
\newblock \href
  {http://dx.doi.org/https://doi.org/10.1016/B978-0-12-385512-1.00003-7}
  {\path{doi:https://doi.org/10.1016/B978-0-12-385512-1.00003-7}}.

\bibitem{7373667}
M.~Ghamkhari, A.~Wierman, H.~Mohsenian-Rad, Energy portfolio optimization of
  data centers, IEEE Transactions on Smart Grid 8~(4) (2017) 1898--1910.

\bibitem{7279063}
M.~Dayarathna, Y.~Wen, R.~Fan, Data center energy consumption modeling: A
  survey, IEEE Communications Surveys Tutorials 18~(1) (2016) 732--794.

\bibitem{BELOGLAZOV2012755}
A.~Beloglazov, J.~Abawajy, R.~Buyya, Energy-aware resource allocation
  heuristics for efficient management of data centers for cloud computing,
  Future Generation Computer Systems 28~(5) (2012) 755 -- 768, special Section:
  Energy efficiency in large-scale distributed systems.
\newblock \href
  {http://dx.doi.org/https://doi.org/10.1016/j.future.2011.04.017}
  {\path{doi:https://doi.org/10.1016/j.future.2011.04.017}}.

\bibitem{Hameed2016}
A.~Hameed, A.~Khoshkbarforoushha, R.~Ranjan, P.~P. Jayaraman, J.~Kolodziej,
  P.~Balaji, S.~Zeadally, Q.~M. Malluhi, N.~Tziritas, A.~Vishnu, S.~U. Khan,
  A.~Zomaya, A survey and taxonomy on energy efficient resource allocation
  techniques for cloud computing systems, Computing 98~(7) (2016) 751--774.

\bibitem{7217791}
M.~A.~A. Faruque, K.~Vatanparvar, Energy management-as-a-service over fog
  computing platform, IEEE Internet of Things Journal 3~(2) (2016) 161--169.

\bibitem{6686006}
I.~Pietri, M.~Malawski, G.~Juve, E.~Deelman, J.~Nabrzyski, R.~Sakellariou,
  Energy-constrained provisioning for scientific workflow ensembles, in: 2013
  International Conference on Cloud and Green Computing, 2013, pp. 34--41.

\bibitem{BAKER201796}
T.~Baker, M.~Asim, H.~Tawfik, B.~Aldawsari, R.~Buyya, An energy-aware service
  composition algorithm for multiple cloud-based iot applications, Journal of
  Network and Computer Applications 89 (2017) 96 -- 108, emerging Services for
  Internet of Things (IoT).
\newblock \href {http://dx.doi.org/https://doi.org/10.1016/j.jnca.2017.03.008}
  {\path{doi:https://doi.org/10.1016/j.jnca.2017.03.008}}.

\bibitem{7448886}
M.~Shojafar, N.~Cordeschi, E.~Baccarelli, Energy-efficient adaptive resource
  management for real-time vehicular cloud services, IEEE Transactions on Cloud
  Computing (2018) 1--1.

\bibitem{Georgiou:2018:YPL:3196398.3196414}
S.~Georgiou, M.~Kechagia, P.~Louridas, D.~Spinellis, What are your programming
  language's energy-delay implications?, in: Proceedings of the 15th
  International Conference on Mining Software Repositories, MSR '18, ACM, New
  York, NY, USA, 2018, pp. 303--313.
\newblock \href {http://dx.doi.org/10.1145/3196398.3196414}
  {\path{doi:10.1145/3196398.3196414}}.

\bibitem{Alan:2015:EDT:2807591.2807628}
I.~Alan, E.~Arslan, T.~Kosar, Energy-aware data transfer algorithms, in:
  Proceedings of the International Conference for High Performance Computing,
  Networking, Storage and Analysis, SC '15, ACM, New York, NY, USA, 2015, pp.
  44:1--44:12.
\newblock \href {http://dx.doi.org/10.1145/2807591.2807628}
  {\path{doi:10.1145/2807591.2807628}}.

\bibitem{Pietri:2018:SDS:3221269.3221298}
I.~Pietri, R.~Sakellariou, Scheduling data-intensive scientific workflows with
  reduced communication, in: Proceedings of the 30th International Conference
  on Scientific and Statistical Database Management, SSDBM '18, ACM, New York,
  NY, USA, 2018, pp. 25:1--25:4.
\newblock \href {http://dx.doi.org/10.1145/3221269.3221298}
  {\path{doi:10.1145/3221269.3221298}}.

\bibitem{lambert2018}
T.~Lambert, R.~Sakellariou, Allocation of publisher/subscriber data links on a
  set of virtual machines, in: 2018 IEEE 11th International Conference on Cloud
  Computing, 2018.

\bibitem{Xu2013}
H.~Xu, B.~Li, Joint request mapping and response routing for geo-distributed
  cloud services, in: 2013 Proceedings IEEE INFOCOM, 2013, pp. 854--862.

\bibitem{Hung2015}
C.-C. Hung, L.~Golubchik, M.~Yu, Scheduling jobs across geo-distributed
  datacenters, in: Proceedings of the Sixth ACM Symposium on Cloud Computing,
  SoCC '15, ACM, New York, NY, USA, 2015, pp. 111--124.
\newblock \href {http://dx.doi.org/10.1145/2806777.2806780}
  {\path{doi:10.1145/2806777.2806780}}.

\bibitem{Heintz2016}
B.~Heintz, A.~Chandra, R.~K. Sitaraman, J.~Weissman, End-to-end optimization
  for geo-distributed mapreduce, IEEE Transactions on Cloud Computing 4~(3)
  (2016) 293--306.

\bibitem{Sakr2011}
S.~Sakr, A.~Liu, D.~M. Batista, M.~Alomari, A survey of large scale data
  management approaches in cloud environments, IEEE Communications Surveys
  Tutorials 13~(3) (2011) 311--336.

\bibitem{Yang2017}
C.~Yang, Q.~Huang, Z.~Li, K.~Liu, F.~Hu, Big data and cloud computing:
  innovation opportunities and challenges, International Journal of Digital
  Earth 10~(1) (2017) 13--53.
\newblock \href
  {http://arxiv.org/abs/https://doi.org/10.1080/17538947.2016.1239771}
  {\path{arXiv:https://doi.org/10.1080/17538947.2016.1239771}}, \href
  {http://dx.doi.org/10.1080/17538947.2016.1239771}
  {\path{doi:10.1080/17538947.2016.1239771}}.

\bibitem{7867735}
Z.~Wen, R.~Yang, P.~Garraghan, T.~Lin, J.~Xu, M.~Rovatsos, Fog orchestration
  for internet of things services, IEEE Internet Computing 21~(2) (2017)
  16--24.

\bibitem{Yi2015}
S.~Yi, Z.~Qin, Q.~Li, Security and privacy issues of fog computing: A survey,
  in: K.~Xu, H.~Zhu (Eds.), Wireless Algorithms, Systems, and Applications,
  Springer International Publishing, Cham, 2015, pp. 685--695.

\bibitem{Dean2008}
J.~Dean, S.~Ghemawat, Mapreduce: Simplified data processing on large clusters,
  Commun. ACM 51~(1) (2008) 107--113.
\newblock \href {http://dx.doi.org/10.1145/1327452.1327492}
  {\path{doi:10.1145/1327452.1327492}}.

\bibitem{Greenberg2008}
A.~Greenberg, J.~Hamilton, D.~A. Maltz, P.~Patel, The cost of a cloud: Research
  problems in data center networks, SIGCOMM Comput. Commun. Rev. 39~(1) (2008)
  68--73.
\newblock \href {http://dx.doi.org/10.1145/1496091.1496103}
  {\path{doi:10.1145/1496091.1496103}}.

\bibitem{Vulimiri2015}
A.~Vulimiri, C.~Curino, P.~B. Godfrey, T.~Jungblut, J.~Padhye, G.~Varghese,
  Global analytics in the face of bandwidth and regulatory constraints, in:
  Proceedings of the 12th USENIX Conference on Networked Systems Design and
  Implementation, NSDI'15, USENIX Association, Berkeley, CA, USA, 2015, pp.
  323--336.

\bibitem{Confais2017}
B.~Confais, A.~Lebre, B.~Parrein, Performance Analysis of Object Store Systems
  in a Fog and Edge Computing Infrastructure, Springer Berlin Heidelberg,
  Berlin, Heidelberg, 2017, pp. 40--79.

\bibitem{Bellavista2017}
P.~Bellavista, A.~Zanni, Feasibility of fog computing deployment based on
  docker containerization over raspberrypi, in: Proceedings of the 18th
  International Conference on Distributed Computing and Networking, ICDCN '17,
  ACM, New York, NY, USA, 2017, pp. 16:1--16:10.
\newblock \href {http://dx.doi.org/10.1145/3007748.3007777}
  {\path{doi:10.1145/3007748.3007777}}.

\bibitem{Vaquero:2014:FYW:2677046.2677052}
L.~M. Vaquero, L.~Rodero-Merino, Finding your way in the fog: Towards a
  comprehensive definition of fog computing, SIGCOMM Comput. Commun. Rev.
  44~(5) (2014) 27--32.
\newblock \href {http://dx.doi.org/10.1145/2677046.2677052}
  {\path{doi:10.1145/2677046.2677052}}.

\bibitem{Velasquez2018}
K.~Velasquez, D.~P. Abreu, M.~R.~M. Assis, C.~Senna, D.~F. Aranha, L.~F.
  Bittencourt, N.~Laranjeiro, M.~Curado, M.~Vieira, E.~Monteiro, E.~Madeira,
  Fog orchestration for the internet of everything: state-of-the-art and
  research challenges, Journal of Internet Services and Applications 9~(1)
  (2018) 14.
\newblock \href {http://dx.doi.org/10.1186/s13174-018-0086-3}
  {\path{doi:10.1186/s13174-018-0086-3}}.

\bibitem{Velasquez2017}
K.~Velasquez, D.~P. Abreu, M.~Curado, E.~Monteiro, Service placement for
  latency reduction in the internet of things, Annals of Telecommunications
  72~(1) (2017) 105--115.
\newblock \href {http://dx.doi.org/10.1007/s12243-016-0524-9}
  {\path{doi:10.1007/s12243-016-0524-9}}.

\bibitem{Skarlat2017}
O.~Skarlat, M.~Nardelli, S.~Schulte, M.~Borkowski, P.~Leitner, Optimized iot
  service placement in the fog, Service Oriented Computing and Applications
  11~(4) (2017) 427--443.
\newblock \href {http://dx.doi.org/10.1007/s11761-017-0219-8}
  {\path{doi:10.1007/s11761-017-0219-8}}.

\bibitem{Ravindra2017}
P.~Ravindra, A.~Khochare, S.~P. Reddy, S.~Sharma, P.~Varshney, Y.~Simmhan,
  Echo: An adaptive orchestration platform for hybrid dataflows across cloud
  and edge, in: M.~Maximilien, A.~Vallecillo, J.~Wang, M.~Oriol (Eds.),
  Service-Oriented Computing, Springer International Publishing, Cham, 2017,
  pp. 395--410.

\bibitem{Kim2018}
N.~Y. Kim, J.~H. Ryu, B.~W. Kwon, Y.~Pan, J.~H. Park, Cf-cloudorch: container
  fog node-based cloud orchestration for iot networks, The Journal of
  Supercomputing\href {http://dx.doi.org/10.1007/s11227-018-2493-4}
  {\path{doi:10.1007/s11227-018-2493-4}}.

\bibitem{8114500}
K.~Velasquez, D.~P. Abreu, D.~Gonçalves, L.~Bittencourt, M.~Curado,
  E.~Monteiro, E.~Madeira, Service orchestration in fog environments, in: 2017
  IEEE 5th International Conference on Future Internet of Things and Cloud
  (FiCloud), 2017, pp. 329--336.

\bibitem{7946419}
M.~S. de~Brito, S.~Hoque, T.~Magedanz, R.~Steinke, A.~Willner, D.~Nehls,
  O.~Keils, F.~Schreiner, A service orchestration architecture for fog-enabled
  infrastructures, in: 2017 Second International Conference on Fog and Mobile
  Edge Computing (FMEC), 2017, pp. 127--132.

\bibitem{santos2017fog}
J.~Santos, T.~Wauters, B.~Volckaert, F.~De~Turck, Fog computing: Enabling the
  management and orchestration of smart city applications in 5g networks,
  Entropy 20~(1) (2017) 4.

\bibitem{Lee2016}
C.~Lee, {Cloud Federation Management and Beyond: Requirements, Relevant
  Standards, and Gaps}, {IEEE Cloud Computing} 3~(1) (2016) pp. 42--49.

\bibitem{NIST:CCTR1}
{NIST}, {NIST US Government Cloud Computing Technology Roadmap, Volume I: High
  Priority Requirements to Further USG Agency Cloud Computing Adoption},
  {Special Publication 500-293} (November 2011).

\bibitem{InCommon}
{InCommon}, {InCommon}, \url{http://incommon.org}.

\bibitem{eduGAIN}
{eduGAIN}, {eduGAIN}, \url{http://www.edugain.org}.

\bibitem{IGTF}
{IGTF}, {The Interoperable Global Trust Federation}, https://www.igtf.net.

\bibitem{GlobusAuth2016}
S.~Tuecke, R.~Ananthakrishnan, K.~Chard, M.~Lidman, B.~McCollam, S.~Rosen,
  I.~Foster, Globus auth: A research identity and access management platform,
  in: 2016 IEEE 12th International Conference on e-Science (e-Science), 2016,
  pp. 203--212.

\bibitem{WG_KO}
J.~Messina, B.~Bohn, S.~Diamond, {NIST Public Working Group on Federated Cloud
  (PWGFC) IEEE P2302 Intercloud Kickoff},
  {http://sites.ieee.org/sagroups-2302/files/2017/08/NIST-PWGFC-IEEE-P2302-Kickoff-31Aug17.pdf}.

\bibitem{NIST_Fed_Ref_Arch}
{NIST}, {The NIST Cloud Federation Reference Architecture},
  \url{https://collaborate.nist.gov/twiki-cloud-computing/bin/view/
  CloudComputing/FederatedCloudPWGFC}.

\bibitem{Stoj2015}
I.~Stojmenovic, S.~Wen, X.~Huang, H.~Luan, An overview of fog computing and its
  security issues, Concurrency and Computation: Practice and Experience 28~(10)
  (2015) 2991--3005.
\newblock \href
  {http://arxiv.org/abs/https://onlinelibrary.wiley.com/doi/pdf/10.1002/cpe.3485}
  {\path{arXiv:https://onlinelibrary.wiley.com/doi/pdf/10.1002/cpe.3485}},
  \href {http://dx.doi.org/10.1002/cpe.3485} {\path{doi:10.1002/cpe.3485}}.

\bibitem{OS_Keystone_Fed_API}
{The OpenStack Foundation}, {Federated Identity},
  \url{https://docs.openstack.org/keystone/pike/admin/
  federated-identity.html}.

\bibitem{CILogon}
{The CILogon Project}, {CILogon: An Integrated Identity and Access Management
  Platform for Science}, \url{https://www.cilogon.org}.

\bibitem{GEANT_TCS}
{The G\'EANT Project}, {TCS - Trusted Certificate Service},
  \url{https://www.geant.org/Services/Trust\_identity\_and\_security/
  Pages/TCS.aspx}.

\bibitem{Zheng2018BlockchainCA}
Z.~Zheng, S.~Xie, H.~Wang, Blockchain challenges and opportunities : A survey,
  2018.

\bibitem{Amadeo2015}
M.~Amadeo, C.~Campolo, A.~Iera, A.~Molinaro, Named data networking for iot: An
  architectural perspective, in: 2014 European Conference on Networks and
  Communications (EuCNC), 2014, pp. 1--5.

\bibitem{Sahai2005}
A.~Sahai, B.~Waters, Fuzzy identity-based encryption, in: Proceedings of the
  24th Annual International Conference on Theory and Applications of
  Cryptographic Techniques, EUROCRYPT'05, Springer-Verlag, Berlin, Heidelberg,
  2005, pp. 457--473.
\newblock \href {http://dx.doi.org/10.1007/11426639\_27}
  {\path{doi:10.1007/11426639\_27}}.

\bibitem{Bethencourt2007}
J.~Bethencourt, A.~Sahai, B.~Waters, Ciphertext-policy attribute-based
  encryption, in: Proceedings of the 2007 IEEE Symposium on Security and
  Privacy, SP '07, IEEE Computer Society, Washington, DC, USA, 2007, pp.
  321--334.
\newblock \href {http://dx.doi.org/10.1109/SP.2007.11}
  {\path{doi:10.1109/SP.2007.11}}.

\bibitem{Yu2015b}
Y.~Yu, A.~Afanasyev, D.~Clark, V.~Jacobson, L.~Zhang, et~al., Schematizing
  trust in named data networking, in: Proceedings of the 2nd International
  Conference on Information-Centric Networking, ACM, 2015, pp. 177--186.

\bibitem{Josang07}
{A. J\o sang and R. Ismail and C. Boyd}, {A Survey of Trust and Reputation
  Systems for Online Service Provision}, {Decision Support Systems} 43 (2007)
  618--644.

\bibitem{zheng2014urban}
Y.~Zheng, L.~Capra, O.~Wolfson, H.~Yang, Urban computing: concepts,
  methodologies, and applications, ACM Transactions on Intelligent Systems and
  Technology (TIST) 5~(3) (2014) 38.

\bibitem{IoT:2014}
L.~D. Xu, W.~He, S.~Li, Internet of things in industries: A survey, IEEE
  Transactions on Industrial Informatics 10~(4) (2014) 2233--2243.

\bibitem{Zheng2011}
Y.~Zheng, Location-based social networks: Users, in: Computing with spatial
  trajectories. Zheng, Yu and Zhou, Xiaofang, Springer press, 2011.

\bibitem{Zheng2012}
Y.~Zheng, {Tutorial on Location-Based Social Networks}, in: Proc.\ of WWW'12,
  Lyon, France, 2012.

\bibitem{traynor2012location}
D.~Traynor, K.~Curran, Location-based social networks, From Government to
  E-Governance: Public Administration in the Digital Age (2012) 243.

\bibitem{chen2016dynamic}
N.~Chen, Y.~Chen, Y.~You, H.~Ling, P.~Liang, R.~Zimmermann, Dynamic urban
  surveillance video stream processing using fog computing, in: 2016 IEEE
  second international conference on multimedia big data (BigMM), IEEE, 2016,
  pp. 105--112.

\bibitem{barbosa2014structured}
L.~Barbosa, K.~Pham, C.~Silva, M.~R. Vieira, J.~Freire, Structured open urban
  data: understanding the landscape, Big data 2~(3) (2014) 144--154.

\bibitem{MucelliRezendeOliveira2017176}
E.~M.~R. Oliveira, A.~C. Viana, K.~Naveen, C.~Sarraute, Mobile data traffic
  modeling: Revealing temporal facets, Computer Networks 112 (2017) 176 -- 193.

\bibitem{naboulsi2014classifying}
D.~Naboulsi, R.~Stanica, M.~Fiore, Classifying call profiles in large-scale
  mobile traffic datasets, in: Proc.\ of INFOCOM'14, IEEE, Toronto, Canada,
  2014, pp. 1806--1814.

\bibitem{FourquareUsers}
Foursquare, {About Us}, Fourquare, \url{https://foursquare.com/about} (2017).

\bibitem{wazeUsers}
K.~Hall-Geisler, {Waze and Esri make app-to-infrastructure possible}, Tech
  Crunch, \url{https://goo.gl/HtJxGH} (2016).

\bibitem{instagramUsers}
A.~Heath, {Instagram's user base has doubled in the last 2 years to 700
  million}, Business Insider, \url{https://goo.gl/PWgLVe} (2017).

\bibitem{twitterUsers}
Twitter, {It's what's happening}, Twitter.com, \url{https://goo.gl/Mn6R4U}
  (2017).

\bibitem{Bonomi2014}
F.~Bonomi, R.~Milito, P.~Natarajan, J.~Zhu, Fog Computing: A Platform for
  Internet of Things and Analytics, Springer International Publishing, Cham,
  2014, pp. 169--186.

\bibitem{tang2015hierarchical}
B.~Tang, Z.~Chen, G.~Hefferman, T.~Wei, H.~He, Q.~Yang, A hierarchical
  distributed fog computing architecture for big data analysis in smart cities,
  in: Proceedings of the ASE BigData \& SocialInformatics 2015, ACM, 2015,
  p.~28.

\bibitem{mukhopadhyay2015wearable}
S.~C. Mukhopadhyay, Wearable sensors for human activity monitoring: A review,
  IEEE sensors journal 15~(3) (2015) 1321--1330.

\bibitem{kiourti2017review}
A.~Kiourti, K.~S. Nikita, A review of in-body biotelemetry devices:
  Implantables, ingestibles, and injectables, IEEE Trans. Biomed. Eng 64~(7)
  (2017) 1422--1430.

\bibitem{kumar2010cloud}
K.~Kumar, Y.-H. Lu, Cloud computing for mobile users: Can offloading
  computation save energy?, Computer 43~(4) (2010) 51--56.

\bibitem{you2017energy}
C.~You, K.~Huang, H.~Chae, B.-H. Kim, Energy-efficient resource allocation for
  mobile-edge computation offloading, IEEE Transactions on Wireless
  Communications 16~(3) (2017) 1397--1411.

\bibitem{kosta2012thinkair}
S.~Kosta, A.~Aucinas, P.~Hui, R.~Mortier, X.~Zhang, Thinkair: Dynamic resource
  allocation and parallel execution in the cloud for mobile code offloading,
  in: Infocom, 2012 Proceedings IEEE, IEEE, 2012, pp. 945--953.

\bibitem{kumar2013survey}
K.~Kumar, J.~Liu, Y.-H. Lu, B.~Bhargava, A survey of computation offloading for
  mobile systems, Mobile Networks and Applications 18~(1) (2013) 129--140.

\bibitem{orsini2015computing}
G.~Orsini, D.~Bade, W.~Lamersdorf, Computing at the mobile edge: Designing
  elastic android applications for computation offloading, in: IFIP Wireless
  and Mobile Networking Conference (WMNC), 2015 8th, IEEE, 2015, pp. 112--119.

\bibitem{7399400}
T.~Taleb, A.~Ksentini, P.~Frangoudis, Follow-me cloud: When cloud services
  follow mobile users, IEEE Transactions on Cloud Computing (2017) 1--1.

\bibitem{mahmud2018fog}
R.~Mahmud, R.~Kotagiri, R.~Buyya, Fog computing: A taxonomy, survey and future
  directions, in: Internet of everything, Springer, 2018, pp. 103--130.

\bibitem{song2010limits}
C.~Song, Z.~Qu, N.~Blumm, A.-L. Barab{\'a}si, Limits of predictability in human
  mobility, Science 327~(5968) (2010) 1018--1021.

\bibitem{diogo2018}
D.~Gonçalves, K.~Velasquez, M.~Curado, L.~F. Bittencourt, E.~Madeira,
  Proactive virtual machine migration in fog environments, in: IEEE Symposium
  on Computers and Communications, IEEE, 2018.

\bibitem{8300317}
D.~Xu, Y.~Li, X.~Chen, J.~Li, P.~Hui, S.~Chen, J.~Crowcroft, A survey of
  opportunistic offloading, IEEE Communications Surveys Tutorials (2018) 1--1.

\bibitem{6714496}
L.~D. Xu, W.~He, S.~Li, Internet of things in industries: A survey, IEEE
  Transactions on Industrial Informatics 10~(4) (2014) 2233--2243.

\bibitem{doi:10.1080/00207543.2017.1308576}
Y.~Liao, F.~Deschamps, E.~de~Freitas Rocha~Loures, L.~F.~P. Ramos, Past,
  present and future of industry 4.0 - a systematic literature review and
  research agenda proposal, International Journal of Production Research
  55~(12) (2017) 3609--3629.
\newblock \href
  {http://arxiv.org/abs/https://doi.org/10.1080/00207543.2017.1308576}
  {\path{arXiv:https://doi.org/10.1080/00207543.2017.1308576}}, \href
  {http://dx.doi.org/10.1080/00207543.2017.1308576}
  {\path{doi:10.1080/00207543.2017.1308576}}.

\bibitem{Kagermann2015}
H.~Kagermann, Change Through Digitization---Value Creation in the Age of
  Industry 4.0, Springer Fachmedien Wiesbaden, Wiesbaden, 2015, pp. 23--45.

\bibitem{Serpanos2018}
D.~Serpanos, M.~Wolf, Industrial Internet of Things, Springer International
  Publishing, Cham, 2018, pp. 37--54.

\bibitem{Jeschke2017}
S.~Jeschke, C.~Brecher, T.~Meisen, D.~{\"O}zdemir, T.~Eschert, Industrial
  Internet of Things and Cyber Manufacturing Systems, Springer International
  Publishing, Cham, 2017, pp. 3--19.

\bibitem{7328144}
V.~Gazis, A.~Leonardi, K.~Mathioudakis, K.~Sasloglou, P.~Kikiras, R.~Sudhaakar,
  Components of fog computing in an industrial internet of things context, in:
  2015 12th Annual IEEE International Conference on Sensing, Communication, and
  Networking - Workshops (SECON Workshops), 2015, pp. 1--6.

\bibitem{HOSSAIN2016192}
M.~S. Hossain, G.~Muhammad, Cloud-assisted industrial internet of things (iiot)
  â€“ enabled framework for health monitoring, Computer Networks 101
  (2016) 192 -- 202, industrial Technologies and Applications for the Internet
  of Things.
\newblock \href
  {http://dx.doi.org/https://doi.org/10.1016/j.comnet.2016.01.009}
  {\path{doi:https://doi.org/10.1016/j.comnet.2016.01.009}}.

\bibitem{Steiner2016}
W.~Steiner, S.~Poledna, Fog computing as enabler for the industrial internet of
  things, e {\&} i Elektrotechnik und Informationstechnik 133~(7) (2016)
  310--314.

\bibitem{b12}
{Industrial Internet Consortium},
  \href{{http://www.iiconsortium.org/IIC\_PUB\_G1\_V1.80\_2017-01-31.pdf}}{{The
  Industrial Internet of Things Reference Architecture}}, {Technical Report},
  {Industrial Internet Consortium} (Jan. 2017).
\newline\urlprefix\url{{http://www.iiconsortium.org/IIC\_PUB\_G1\_V1.80\_2017-01-31.pdf}}

\bibitem{7785890}
K.~Wang, Y.~Wang, Y.~Sun, S.~Guo, J.~Wu, Green industrial internet of things
  architecture: An energy-efficient perspective, IEEE Communications Magazine
  54~(12) (2016) 48--54.

\bibitem{7467436}
J.~Wan, S.~Tang, Z.~Shu, D.~Li, S.~Wang, M.~Imran, A.~V. Vasilakos,
  Software-defined industrial internet of things in the context of industry
  4.0, IEEE Sensors Journal 16~(20) (2016) 7373--7380.

\bibitem{smartcomp}
E.~Kavakli, J.~Buenabad-Ch\'{a}vez, V.~Tountopoulos, P.~Loucopoulos,
  R.~Sakellariou, An architecture for disruption management in smart
  manufacturing, in: {4th IEEE International Conference on Smart Computing
  (SMARTCOMP'18)}, 2018.

\bibitem{es2018}
E.~Kavakli, J.~Buenabad-Ch\'{a}vez, V.~Tountopoulos, P.~Loucopoulos,
  R.~Sakellariou, Specification of a software architecture for an industry 4.0
  environment, in: {The 6th International Conference on Enterprise Systems
  (ES2018)}, 2018.

\bibitem{10.1007/978-981-13-0311-1_30}
M.~Shin, J.~Woo, I.~Wane, S.~Kim, H.-S. Yu, Implementation of security
  mechanism in iiot systems, in: S.~O. Hwang, S.~Y. Tan, F.~Bien (Eds.),
  Proceedings of the Sixth International Conference on Green and Human
  Information Technology, Springer Singapore, Singapore, 2019, pp. 183--187.

\bibitem{7445139}
A.~Sajid, H.~Abbas, K.~Saleem, Cloud-assisted iot-based scada systems security:
  A review of the state of the art and future challenges, IEEE Access 4 (2016)
  1375--1384.

\bibitem{7883994}
M.~Wollschlaeger, T.~Sauter, J.~Jasperneite, The future of industrial
  communication: Automation networks in the era of the internet of things and
  industry 4.0, IEEE Industrial Electronics Magazine 11~(1) (2017) 17--27.

\bibitem{MOURTZIS2016290}
D.~Mourtzis, E.~Vlachou, N.~Milas, Industrial big data as a result of iot
  adoption in manufacturing, Procedia CIRP 55 (2016) 290 -- 295, 5th CIRP
  Global Web Conference - Research and Innovation for Future Production (CIRPe
  2016).
\newblock \href
  {http://dx.doi.org/https://doi.org/10.1016/j.procir.2016.07.038}
  {\path{doi:https://doi.org/10.1016/j.procir.2016.07.038}}.

\bibitem{TAO2018}
F.~Tao, Q.~Qi, A.~Liu, A.~Kusiak, Data-driven smart manufacturing, Journal of
  Manufacturing Systems\href
  {http://dx.doi.org/https://doi.org/10.1016/j.jmsy.2018.01.006}
  {\path{doi:https://doi.org/10.1016/j.jmsy.2018.01.006}}.

\bibitem{8259028}
J.~Fu, Y.~Liu, H.~Chao, B.~Bhargava, Z.~Zhang, Secure data storage and
  searching for industrial iot by integrating fog computing and cloud
  computing, IEEE Transactions on Industrial Informatics (2018) 1--1.

\bibitem{es2018b}
V.~Tountopoulos, E.~Kavakli, R.~Sakellariou, Towards a cloud-based controller
  for data-driven service orchestration in smart manufacturing, in: {The 6th
  International Conference on Enterprise Systems (ES2018)}, 2018.

\bibitem{alma9956512960001631}
Q.~Zhang, Q.~Zhang, L.~T. Yang, Z.~Chen, P.~Li, F.~Bu, An adaptive droupout
  deep computation model for industrial iot big data learning with
  crowdsourcing to cloud computing, IEEE transactions on industrial
  informatics. (2018) 1,1.

\bibitem{7428027}
H.~Jayakumar, A.~Raha, Y.~Kim, S.~Sutar, W.~S. Lee, V.~Raghunathan,
  Energy-efficient system design for iot devices, in: 2016 21st Asia and South
  Pacific Design Automation Conference (ASP-DAC), 2016, pp. 298--301.

\bibitem{samos2018}
R.~Sakellariou, J.~Buenabad-Ch\'avez, E.~Kavakli, I.~Spais, V.~Tountopoulos,
  {High Performance Computing and Industry 4.0: Experiences from the DISRUPT
  Project}, in: 2018 International Conference on Embedded Computer Systems:
  Architectures, Modeling, and Simulation (SAMOS XVIII), 2018.

\bibitem{8029323}
W.~A. Higashino, M.~A.~M. Capretz, L.~F. Bittencourt, Cepaas: Complex event
  processing as a service, in: 2017 IEEE International Congress on Big Data
  (BigData Congress), 2017, pp. 169--176.

\bibitem{7167238}
A.~Sadeghi, C.~Wachsmann, M.~Waidner, Security and privacy challenges in
  industrial internet of things, in: 2015 52nd ACM/EDAC/IEEE Design Automation
  Conference (DAC), 2015, pp. 1--6.

\end{thebibliography}

\end{document}